\documentclass[prb,twocolumn,superscriptaddress,showpacs,amsmath,amssymb,aps]{revtex4-1}
 \usepackage{graphicx}
\usepackage{amsmath}
\usepackage{amssymb}
\usepackage{float}
\usepackage{color}

\usepackage[T1]{fontenc}
\usepackage{array}
\usepackage{booktabs}

\usepackage{color}
\usepackage{soul,xcolor}
\usepackage{ulem}
\setstcolor{red}

\usepackage[columnwise]{lineno}

\begin{document}
%\title{Spin qubit shuttling between coupled quantum dots with   anisotropic Land\'{e} g-factors   }
\title{ Spin qubit  shuttling between coupled quantum dots with  inhomogeneous Land\'{e}  g-tensors  }

%%%%%%%%%%%%%%%%%%%%%%%%%%%%%%%%%%%%%%%%%%%%%%%%%
%%%%%%%%%%%%%%%%%%%%%%%%%%%%%%%%%%%%%%%%%%%%%%%%%

\author{Zhi-Hai\!  Liu }
\email{liuzh@baqis.ac.cn}
\affiliation{Beijing Academy of Quantum Information Sciences, Beijing 100193, China}

 \author{Xiao-Fei\! Liu}
 \email{liuxf@baqis.ac.cn}
 \affiliation{Beijing Academy of Quantum Information Sciences, Beijing 100193, China}

 \author{H. Q. Xu}
\email{hqxu@pku.edu.cn}
 \affiliation{Beijing Academy of Quantum Information Sciences, Beijing 100193, China}
  \affiliation{ Beijing Key Laboratory of Quantum Devices,  Peking University, Beijing 100871, China}

\begin{abstract}
 By utilizing   the site-dependent spin quantization axis in  semiconductor  quantum dot (QD) arrays,  shuttling-based spin qubit gates have become an appealing approach to realize scalable quantum computing
 due to the circumvention of using high-frequency driving fields.  The emergence of a   spin deviation  from   the local   quantization axis  of one residing  QD  is   the  prerequisite     to implement the  qubit gates. In this work, we study the non-adiabatic dynamics of a spin qubit shuttling between coupled QDs with    inhomogeneous Land\'{e}  g-tensors  and  a small   magnetic field.  The  spin dynamics  is analyzed  through solving the time-dependent Schr\"{o}dinger equation  of the qubit under the effects of spin-orbit interaction and   rapid ramping inter-dot detuning.  The precondition,  imposed on   the ramping  time and the tunnel-coupling strength,  to ensure a high-fidelity  inter-dot transfer is  estimated.   We then calculate the change in the spin orientation  of a  transferred  qubit, and study the dependences of the spin deviation  on the difference in the  quantization axes of the two QDs, the tunnel-coupling strength, and the ramping time. We also demonstrate that the effect of multiple rounds of inter-dot bidirectional shuttling can be captured by an operator matrix, and evaluate the idling times required for realizing the single-qubit Pauli-X and  Pauli-Y gates. Intriguingly, it is confirmed that a generalized Hadamard gate can be achieved through tuning the idling times.
 \end{abstract}
%orientation change
%%%%%%%%%%%%%%%%%%%%%%%%%%%%%%%%%%%%%%%%%%%%%%%%%
%%%%%%%%%%%%%%%%%%%%%%%%%%%%%%%%%%%%%%%%%%%%%%%%%

\date{\today}
\maketitle
%%%%%%%%%%%%%%%%%%%%%%%%%%%%%%%%%%%%%%%%%%%%%%%%%
%%%%%%%%%%%%%%%%%%%%%%%%%%%%%%%%%%%%%%%%%%%%%%%%% ~\cite{Xue2024,Mills2019,Noiri2022,Smet2025,Noiri2022,Unseld2025,Wang2024,John2024,Pazhedath2025,Langrock2023,Jeon2025,Zwerver2023}

 \section{Introduction}
 Semiconductor quantum dots (QDs) have attracted extensive attentions owing to  their  great potential for  implementing universal  quantum   computation~\cite{Burkard2023,Chatterjee2021,Stano2022,Weinstein2023}.    Experimentally, the single- and two-qubit gate fidelities  of spin qubits in silicon-based~\cite{Noiri2022a,Mills2022,Xue2022,Philips2022,Steinacker2025} and Ge-based~\cite{Lawrie2023,Hendrickx2021,Hendrickx2020}  QDs have exceeded the quantum error correction threshold, thereby paving the way for the practical implementation of noisy intermediate-scale quantum devices~\cite{Warren2022}.  Most recently, the engineering of  spin qubit shuttling     in dense QD arrays has reduced the need for high-frequency oscillating drivers in   qubit manipulation~\cite{Xue2024,Smet2025,Langrock2023,Jeon2025,Zwerver2023,Fer2024,Noiri2022,Unseld2025,Wang2024,John2024}.  To elaborate, conveyor-mode shuttling facilitates high-fidelity state transfer over extended distances~\cite{Xue2024,Smet2025,Langrock2023,Jeon2025,Zwerver2023}, whereas  coherent spin shuttling between QDs enables the realization of  qubit quantum logic gates~\cite{Fer2024,Noiri2022,Unseld2025,Wang2024,John2024}.  Importantly, these techniques also alleviate crosstalk and heating effects in large-scale scalable quantum computing~\cite{Kunne2024,Seidler2022,Sato2025,Siegel2024,Crawford2023,Bosco2024}.%

Essentially, the shuttling-based single-qubit logic gates
 in Refs.~\onlinecite{Unseld2025,Wang2024,John2024}  are established on the Larmor precession
 of a hole (or electron) spin qubit transferred into one target QD.
 The emergence of   a  deviation   from  the local spin quantization axis  is  the basic  premise  for  implementing the  qubit gates~\cite{Doelman2024}.
This evidently requires a  non-adiabatic  transfer process, otherwise  the spin direction  of a shuttled  qubit is  bound to be  aligned with the local axis~\cite{Liu2024,Kandel2021,Fujita2017}. Remarkably,   it goes against  to  the suppression of   Landau-Zener transitions (LZTs),  which leads to the reduction of the fidelity of an inter-dot transfer~\cite{Petta2010,Cao2013,Ota2018,Ginzel2020,Lucak2017}.  Equipped with precise control of the inter-dot detuning, the   logic gates are actually implemented in an operational regime that not only enables high-fidelity qubit transfer but also ensures a well-defined spin deviation. In contrast to studies on  LZT processes \cite{Ota2018,Ginzel2020,Lima2025,Lucak2017}, however, the spin deviation of qubits transferred to the target  QD has thus far received insufficient theoretical attention. Accordingly, there is an urgent need to thoroughly analyze the evolution of qubit spin orientation following non-adiabatic inter-dot shuttling processes.

Considering a non-orthogonal deviation of the transferred qubit from the local spin-quantization axis, Pauli-X or Pauli-Y qubit gates are generally implemented via multiple rounds of inter-dot transfer ~\cite{Unseld2025,Wang2024,John2024}.
   Interestingly, Ref.~\onlinecite{Li2018a} reported an alternative theoretical approach for realizing qubit gates along unidirectional simultaneous transport in double  QDs. Therein, quaternionic algebra has thus far been harnessed for the inverse design of single-qubit phase and NOT gates, in accordance with the  system's Hamiltonian structure~\cite{Li2018a,Li2018b}. By comparison,   state-of-the-art experimental logic protocols leveraging multiple-round  transfer driven by inter-dot detuning exhibit substantial flexibility~\cite{Unseld2025,Wang2024,John2024}, albeit at the expense of prolonged gate operation times.   Particularly, fine-tuning the idling times within the    QDs  constitutes an indispensable prerequisite for implementing shuttling-based single-qubit gates ~\cite{Unseld2025,Wang2024,John2024}. Hitherto, to the best of  our knowledge, the  dependence  of the requisite times on the   characteristics of   inter-dot  shuttling
 still remains to be elucidated for
  pursuing high-fidelity and  more sophisticated  shuttling-based qubit gates.

In this work,  we focus on    the non-adiabatic  dynamics of a spin qubit shuttling between two  coupled QDs  with inhomogeneous    Land\'{e} g-tensors~\cite{Mu2021,Zhang2021,Hendrickx2024,Bree2016}, as such the spin quantization axis is site-dependent in a  small magnetic field.   Firstly, we  evaluate the  infidelity of an inter-dot  transfer of the qubit  under the effects of spin-orbit interaction (SOI)~\cite{Fer2024,Bosco2024} and the rapid ramping inter-dot detuning~\cite{Petta2010,Doelman2024}.  The precondition, imposed on the ramping time and the tunnel-coupling strength, is estimated to reduce the infidelity of the inter-dot transfer.    Following the realization  of a  high-fidelity ($>99\%$) transfer, we subsequently  study the change  in the spin orientation of a transferred   qubit, thereby revealing the  flexibility of deviation from the spin-quantization axis of the target QD.  Besides, we show that such   deviation can be reflected by the  reversal of  the spin orientation of a qubit undergoing one  round of inter-dot transfer.  By introducing an operation matrix   describing  the overall effect of  multiple rounds of  inter-dot shuttling protocols,    the requisite idling times are further evaluated in the implementation of shuttling-based single-qubit  gates. In addition to the realization of Pauli-X and Pauli-Y gates, we demonstrate that a generalized Hadamard gate can also be attained  with  tuning the idling times.

 %

%%%%%%%%%%%%%%%%%%%%%%%%%%%%%%%%%%%%%%%%%%%%%%%%%
%%%%%%%%%%%%%%%%%%%%%%%%%%%%%%%%%%%%%%%%%%%%%%%%%

%%%%%%%%%%%%%%%%%%%%%%%%%%%%%%%%%%%%%%%%%%%%%%%%%
%%%%%%%%%%%%%%%%%%%%%%%%%%%%%%%%%%%%%%%%%%%%%%%%%
%epitaxially
\begin{figure}
  \centering
  \includegraphics[width=0.46\textwidth]{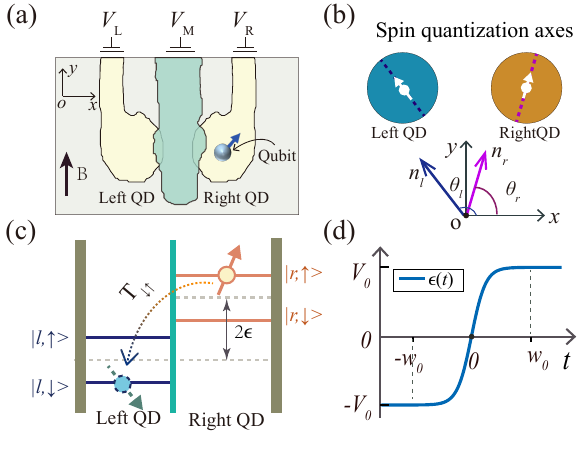}\\
  \caption{(a) Schematic  of   coupled quantum dots, i.e.,  left QD and right QD,  occupied by a single   spin  qubit, in which  $V^{}_{L/R}$ and $V^{}_{M}$ represent the left/right   plunger-gate   and   middle barrier-gate potentials.  (b)  The  spin quantization axes of the two QDs  in    a low magnetic   field $B$ applied in the  $y$-direction, i.e., $\mathbf{n}^{}_{l/r} = (\cos\theta^{}_{l/r}, \sin\theta^{}_{l/r},0 )$ with $\theta^{}_{l/r}$ being the  direction angle  of  the left/right dot.  (c) The ``spin-flipped '' tunneling  $T^{}_{\downarrow \uparrow}$  of a qubit between  the two QDs, with $|l/r,\uparrow\rangle$ and $|l/r,\downarrow\rangle$  indicating the    local  Zeeman-splitting states   and   $ \epsilon$ sizing the inter-dot detuning. (d) A  specific   ramping   up  of    $\epsilon(t)=  V^{}_{0}\tanh(\pi t/w^{}_{0}) $ to implement a leftward inter-dot transfer, with $w^{}_{0}$ being the ramping time.
  % Here,   $V^{}_{r}$ and $V^{}_{l}$ indicate the plunger-gate potentials
}\label{Fig1}
\end{figure}

%the  particle can  move      between the left and right  QDs   through   modulating the interdot  detuning $V^{}_{0}=(V^{}_{r}-V^{}_{l})/2$,
 \section{  The effective  Hamiltonian of a spin qubit in  double QDs}
  The coupled   QDs  we consider   is occupied by  a single  hole (or electron) spin qubit and with  a small   magnetic field $B$ applied in the $y$ direction,  just as shown in Fig.~\ref{Fig1}(a). Considering  the  inhomogeneity in the    Land\'{e}  g-tensors of  the QDs,     the qubit spin-quantization axis   is thereby site-dependent under the transverse magnetic field.    For simplicity,  the   quantization   axes  can  be assumed to be  laid   in the $x$-$y$ plane,      with  $\hat{n}^ {}_{l/r} = (\cos\theta^{}_{ l/r}, \sin \theta^{}_{l/r},0 )$   characterizing the   axis  direction of the left/right QD and with  $\theta^{}_{l}\neq \theta^{}_{r}$ [see Fig.~\ref{Fig1}(b)].
     Similar to self-assembled QDs,     inhomogeneity in the transverse components of   hole Land\'{e}  g-tensors   can be attributed to   fluctuations in  shape symmetry, strain gradients, and structural deformations of the  low-dimensional nanostructures~\cite{Serov2025,Pikus1994,Trifonov2021}. For the gate-defined QDs  focused on in this work,  the transverse inhomogeneous features  can also  be  regulated by the respective gate potentials~\cite{Hendrickx2024} [e.g., $V^{}_{L,R}$  and  $V^{}_{M}$ in Fig.~1(a)].

Notably, when  hyperfine interactions with nuclear spins are considered, random hyperfine interactions can also induce differences in the spin-quantization axes of two coupled QDs~\cite{Oleg2006,Coish2005,Schuetz2014,Glazov2018,Shumilin2015,Mantsevich2019}, in either the absence or presence of a weak external magnetic field. Meanwhile, nuclear fluctuations will disrupt the deterministic distinction between the spin axes (a prerequisite for shuttling-based qubit gates) and exert deleterious effects on the qubit spin coherence~\cite{Glazov2018,Shumilin2015,Mantsevich2019}. To mitigate such adverse impacts, a lower limit is thus imposed on the magnetic-field strength, specifically requiring  $B$ to exceed the magnitude of the averaged hyperfine fields~\cite{Hung2013,Fischer2010,Testelin2009}. Besides, the constraints on the field strength for defining  well-behaved spin qubits in QDs are detailed in Appendix~\ref{S-1}. %Additionally, an upper limit on $B$ is dictated by the need to preserve long coherence times, given the inverse  dependence of the phonon-induced spin relaxation rate on  the magnetic-field magnitude~\cite{Wang2021,Sarkar2023,Golovach2004,Maier2013}.

Based on  the orientation of the local  spin-quantization  axis $\hat{n}^{}_{\kappa=l,r}$, the  low-energy   Zeeman splitting  states of the QDs
   satisfy the relation $(\boldsymbol{\sigma}\cdot \hat{n}^{}_{\kappa})| \kappa,\uparrow/\downarrow\rangle= \pm  | \kappa,\uparrow/\downarrow\rangle$,  where $|\kappa,\uparrow/\downarrow \rangle$ denotes the respective upper/lower  Zeeman-splitting  levels and $\boldsymbol{\sigma}=\{\sigma^{}_{x},\sigma^{}_{y},\sigma^{}_{z}\}^{\rm T}_{}$ represents the  spin Pauli matrices.   The corresponding energy splitting  is denoted by  $\Delta^{}_{z,\kappa}$. Besides,  a mismatch between  the Zeeman splitting  of  two coupled QDs  can give rise to a    spin bottleneck during  resonant tunneling~\cite{Huang2010}.  Accordingly,  we assume   that the energy splitting  is site-independent,   i.e., $\Delta^{}_{ z, \kappa} \equiv \Delta^{}_{z}$,   to pinpoint the  impact of  disparities in the spin-quantization axes of the QDs.

Analogous to the effect of Rashba  SOI~\cite{Geyer2024,Danon2009,Li2014,Liu2018b,Bosco2021},  a difference in the QDs' spin-quantization axes, i.e., $\theta^{}_{l}\neq \theta^{}_{r}$ will lead to the breaking of  spin-conservation rules during the inter-dot
 tunneling.    Then,    the  effective    Hamiltonian   describing  the single spin qubit in  the coupled  QDs  can be written as
\begin{align}
H^{}_{\rm 0} =  &  \sum^{}_{\sigma =\uparrow,\downarrow }  (T^{}_{\sigma^{}_{} \sigma}
    | l,\sigma^{ }_{} \rangle \langle  r,\sigma^{ }_{} |  +T^{}_{\bar{\sigma}\sigma}   | l,\bar{\sigma} ^{ }_{} \rangle \langle  r,\sigma^{ }_{} |+{\rm h.c.} )\nonumber\\
     & +\sum^{}_{\kappa,\sigma }  E^{}_{\kappa,\sigma} | \kappa,\sigma^{ }_{}  \rangle \langle  \kappa,\sigma^{ }_{} |
\ .
\label{H0}
\end{align}
Here,
    $E^{}_{l,\uparrow/\downarrow} =- \epsilon^{}_{}\pm \Delta^{}_{z }/2$ and $E^{}_{r,\uparrow/\downarrow} = \epsilon^{}_{}\pm  \Delta^{}_{z }/2$   indicate the  spin-dependent  on-site energies,  with $\epsilon$ specifying the inter-dot detuning and determined by the plunger-gate  potentials $V^{}_{L,R}$ in Fig.~\ref{Fig1}(a).   $T^{}_{\sigma\sigma}$ and $T_{\bar{\sigma}\sigma}^{}$ quantify  the amplitudes for  a qubit tunneling from the right-QD state $|r,\sigma\rangle$ to the respective spin states $|l,\sigma\rangle $  and $|l,\bar{\sigma}\rangle$ of the other QD. Nominally,   they  can be  referred  to as  the ``spin-conserved'' and ``spin-flipped'' tunneling amplitudes.
 Then,  by  incorporating the effect   of    SOI   in the  inter-dot tunneling processes~\cite{Li2014,Liu2018b}, the  amplitudes  can be analytically derived as
   \begin{align}
 T^{}_{\uparrow\uparrow/\downarrow\downarrow} =  &T^{}_{0}  \left(\cos\varphi^{}_{\rm so} \cos \theta^{}_{\delta} \pm  i \sin\varphi^{}_{\rm so}\sin   \theta^{}_{0} \right)\ ,\nonumber\\
 T^{}_{\uparrow\downarrow/\downarrow\uparrow}= & T^{}_{0}\left(i \cos\varphi_{\rm so} \sin  \theta^{}_{\delta} \pm \sin\varphi^{}_{\rm so}\cos  \theta^{}_{0} \right)\ ,\label{SPT}
  \end{align} in which  $T^{}_{0}$  indicates the   tunnel-coupling  strength dominated by  the barrier-gate  potential $V^{}_{M}$ in Fig.~\ref{Fig1}(a), $ \theta^{}_{0/\delta}=  (\theta^{}_{r} \pm \theta^{}_{l})/2 $, and
 $\varphi^{}_{\rm so} $
 captures the concomitant  effect of SOI during the inter-dot tunneling.    In fact, it depicts the spin rotation phase, named as the Aharonov-Casher (AC) phase, acquired by a  transferred qubit under  the interplay between the SOI and an external magnetic field~\cite{Shekhter2022,Liu2021,Ora2019}. Specifically,  the AC phase  can be  evaluated as  $\varphi^{}_{\rm so} =2d/\ell^{}_{\rm so }$, in which $2d$ characterizes the  inter-dot distance and $\ell^{}_{\rm so}$ indicates the spin-orbit length~\cite{Liu2021}.   Please refer to Appendix~\ref{S-1} for the  details.
Clearly,  the presence of a nonzero AC phase and with $\varphi^{}_{\rm so}\neq n\pi$ ($n=1,2,3,...$) can also lead to  to  the  emergence of ``spin-flipped'' tunneling [see  Fig.~\ref{Fig1}(c)],  even without disparities in the QDs' spin-quantization axes, i.e., $\theta^{}_{\delta}=0$.

%%%%%%%%%%%%%%%%%%%%%%%%%%%%%%%%%%%%%%%%%%%%%%%%%%%%%%%%%%%%%%%%%%%%%%%%%%%%%%%%%%%%%%%%%%%%%%%%%%%%%%%%%%%%%%%%%%%%%%%%%%%%%%%%%%%%%%%%%%%%%%%%%
%%%%%%%%%%%%%%%%%%%%%%%%%%%%%%%%%%%%%%%%%%%%%%%%%%%%%%%%%%%%%%%%%%%%%%%%%%%%%%%%%%%%%%%%%%%%%%%%%%%%%%%%%%%%%%%%%%%%%%%%%%%%%%%%%%%%%%%%%%%%%%%%%

  \section{Non-adiabatic  dynamics of spin qubit shuttling between QDs}

\begin{figure}[tbp!]
  \centering
  \includegraphics[width=0.44\textwidth]{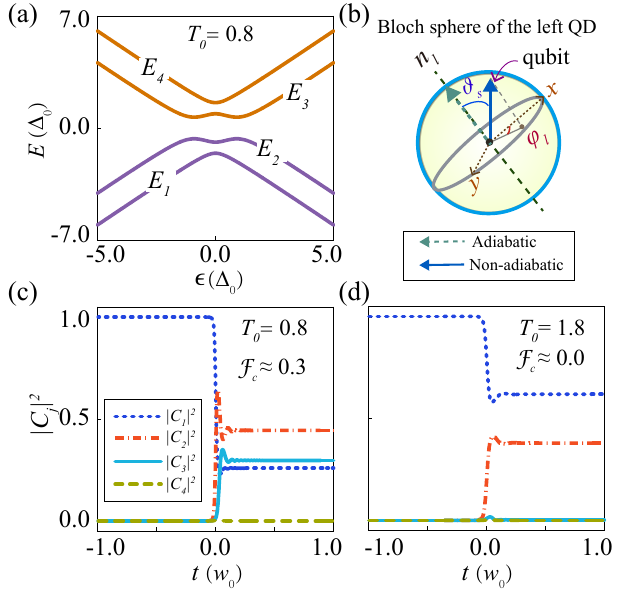}\\
  \caption{ (a)  Energy spectrum of the single qubit as a function of  $\epsilon$  with $\theta^{}_{r}= 0.46\pi$,  $\theta^{}_{l}= 1.11\pi$, $\varphi^{}_{\rm so} =0.12\pi$, $T^{}_{0}=0.8$,  and $\Delta^{}_{0}$ being the  unit of  energy. (b) The spin  orientation of a leftward transferred qubit defined in the Bloch sphere of the left QD  with  $\hat{n}^{}_{l }$ specifying the pole axis  and  the  dashed/solid  arrow  indicating the spin direction  after undergoing a adiabatic/non-adiabatic inter-dot transfer.  (c) and  (d) The time evolutions   of    $|C^{}_{j=1-4} (t)|^{2}_{}$  for a qubit preset in the  ground-state ($E_1$) energy level    with  different $T^{}_{0}$  and    $\epsilon = V^{}_{0}\tanh(\pi t/w^{}_{0})$,  in which $V^{}_{0}= 10$, $w^{}_{0} = 6  t^{}_{0}$, and  $t^{}_{0}= \pi\Delta^{}_{0}/\hbar $.  In addition,       the infidelity of the leftward inter-dot transfer ${\cal F}^{}_{c}$ is indicated  in each case.      %defined in the Bloch sphere of the target (left) QD.
  % Here,   $V^{}_{r}$ and $V^{}_{l}$ indicate the plunger-gate potentials
}\label{Fig2}
\end{figure}

  Enabled by the inter-dot tunnelings in Eqs.~(\ref{SPT}),  a spin qubit  can    move  between the  QDs along  a substantial  change in the detuning~\cite{DiCarlo2004,Zhao2022,Luo2024}.  In order to  realize a leftward inter-dot transfer,  the detuning   $\epsilon$ is expediently  modulated as  $\epsilon(t) =  V^{}_{0}\tanh( \pi t/w^{}_{0})$,    with  $V^{}_{0} \gg \max\{ \Delta^{}_{z },T^{}_{0}\}$ and  $w^{}_{0}$  being the ramping time [see  Fig.~\ref{Fig1}(d)].    The general results we will obtain below   definitely  can be extended to  the scenario of  other rising profiles~\cite{Ginzel2020}.
  Concretely,  Fig.~\ref{Fig2}(a)   show the energy spectrum of the single qubit as a function of $\epsilon$, with   $\Delta^{}_{0} = \Delta^{}_{z } /2$  being the unit of energy.  The  four energies of  $H^{}_{0}(\epsilon) $  are represented by  $E^{}_{j=1-4} (\epsilon)$,   and with   $| \Psi^{}_{j } \rangle  $  indicating the   energy states.

  Theoretically, an  adiabatic transfer  corresponds to  the  case  of  $w^{}_{0} \rightarrow  \infty$, and this can be reflected by the   variations  of   $|\Psi^{}_{1,2}\rangle$ versus $\epsilon$.   Expanded in the normalized  local basis  of $\{e^{i\phi^{}_{0}}_{} |l,\uparrow\rangle,e^{-i\phi^{}_{0}}_{}|l,\downarrow\rangle,
  e^{i\phi^{}_{\delta}}_{}|r,\uparrow\rangle,e^{-i\phi^{}_{\delta}}_{}|r,\downarrow\rangle \}^{}_{}$ with  $\phi^{}_{0/\delta}= [\arg(T^{}_{\uparrow\downarrow})\pm \arg(T^{}_{\uparrow\uparrow})]/2$,  the two   lower-energy states    can be presented as
  %\begin{small}
 \begin{align}
 |\Psi^{}_{1}\rangle=\left(\begin{array}{c}
  \sin\alpha \cos \chi^{}_{1} \\
  \cos\alpha \sin\chi^{}_{1}
\\
 \cos\alpha\cos\chi^{}_{1}\\
- \sin\alpha\sin\chi^{}_{1} \end{array}\right)\ , |\Psi^{}_{2} \rangle =\left(\begin{array}{c}
-   \cos\alpha \sin \chi^{}_{2} \\
  \sin\alpha \cos\chi^{}_{2}
\\
 \sin\alpha\sin\chi^{}_{2}\\
 \cos\alpha\cos\chi^{}_{2} \end{array}\right)\ ,\label{psl}
    \end{align}
   % \end{small}
 in which  $\chi^{}_{n} =[2\arccos(T^{}_{0}\sin\beta/|E^{}_{n}|)+\pi]/4$ and $ \alpha =  \arccos(\epsilon/\Gamma^{}_{0})/2 $ , with   $\Gamma^{}_{0}= \sqrt{\epsilon^{2}_{} +T^{2}_{0}\cos^{2}_{}\beta}$,   and $\beta=\arccos(|T^{}_{\sigma\sigma}|/T^{}_{0})$. Besides,       the   lower  energies  are  calculated as  $ E^{}_{1/2}(\epsilon)=  - \sqrt{ (\Delta^{}_{0} \pm\Gamma^{}_{0})^{2}_{} +T^{2}_{0}\sin^{2}\beta}$.
 On the basis  of $\alpha,\chi^{}_{n }\doteq\pi/2$ at the initial stage of  $\epsilon =  -V^{}_{0}$,   a  qubit   preset in the  spin state $| r,\downarrow/\uparrow  \rangle$   of the right QD is actually kept in the $|\Psi^{}_{1/2 }\rangle$   energy state.   As   $\epsilon$   increases  to $V^{}_{0}$,   the parameter  $\alpha$   changes to   $0$,  and   Eq.~(\ref{psl}) implies that the qubit  will transfer  to  the   state $| l,\downarrow/\uparrow \rangle$  of the other QD. Thus, in this case,  the   spin orientation  of a transferred  qubit is  aligned  with the    spin quantization   axis ($\hat{n}^{}_{l}$) of the   target QD, just as shown in Fig.~\ref{Fig2}(b).

%%%%%%%%%%%%%%%%%%%%%%%%%%%%%%%%%%%%%%%%%%%%%%%%%
%%%%%%%%%%%%%%%%%%%%%%%%%%%%%%%%%%%%%%%%%%%%%%%%%

% the time evolution of

However, within the non-adiabatic regime characterized by a finite $w^{}_{0}$,  transitions between  different  energy  levels  of $H^{}_{0}(\epsilon)$   will complicate   the   inter-dot transfer.   Definitely, at a  fixed time $t$ (of a constant $\epsilon$),  the  evolving state of the qubit  can be expanded  as $\Phi(t)= \sum^{4}_{j=1}C^{}_{j}(t) |\Psi^{}_{j}(t)\rangle$.  The time  evolution  of   the combination coefficient   $C^{}_{j }(t)$  is governed by the  time-dependent Schr\"{o}dinger equation  $i\hbar \partial^{}_{t}  \Phi^{}_{}(t)     =H^{}_{0}(\epsilon)  \Phi^{}_{}(t) $, which can be expanded as
 \begin{align}
i\frac{d C^{}_{j}(t)}{dt}+i\sum^{4}_{j^{\prime}_{}=1}{\cal T}^{}_{j, j^{\prime}_{}}(t)C^{}_{j^{\prime}_{}}(t) =\frac{E^{}_{j}(t)}{\hbar}C^{}_{j}(t)\ ,
    \label{dsf}
\end{align}
with $\hbar$ being the Planck constant and ${\cal T}^{}_{j, j^{\prime}_{}}(t)= \langle\Psi^{}_{j} (t)|\partial^{}_{ t} | \Psi^{}_{j^{\prime}_{}}(t)\rangle $ characterizing the  temporal transition elements.  Based on Eq.~(\ref{psl}), it is found that the  diagonal elements  ${\cal T}^{}_{j,j^{}_{}} =0$, and  the  specific  forms of the off-diagonal elements  ${\cal T}^{}_{ j,j^{\prime}_{}} $ with $j\neq j^{\prime}_{}$ can refer to Appendix~\ref{S-2}.    For a qubit initialized in the  ground-state energy   level at $t=-w_0$, which means  $C^{}_{1} (-w_0)=1$  and $C^{}_{ 2-4}(-w_0)=0$,  Figs.~\ref{Fig2}(c) and \ref{Fig2}(d) exhibit the specific time evolutions of  $|C^{}_{j}(t)|^{2}_{}$    with  two different $T^{}_{0}$.  Because ${\cal T}^{}_{ j,j^{\prime}_{}}(t)\propto  {\rm sech} ^{2}_{} ( \pi t /w^{}_{0})$,  it is evident that
 $|C^{}_{j=1-3}(t)|^{2}_{}$ subject to a rapid change   around  $t=0$, and the qubit will  evolve  into a stable state  as $t$ increases to $w^{}_{0} $.   In addition to the transition between the two low-energy $E^{}_{1,2}$ levels,   the  considerable  magnitude of $ |C^{}_{3}(w^{}_{0})|^{2}_{}$  observed in  Fig.~\ref{Fig2}(c) for a smaller  $T^{}_{0}$  indicates a high probability of  the qubit transition to the   $E_{3}$  level.

% In addition to the transition between the two low-energy levels,

\begin{figure}
  \centering
  \includegraphics[width=0.49\textwidth]{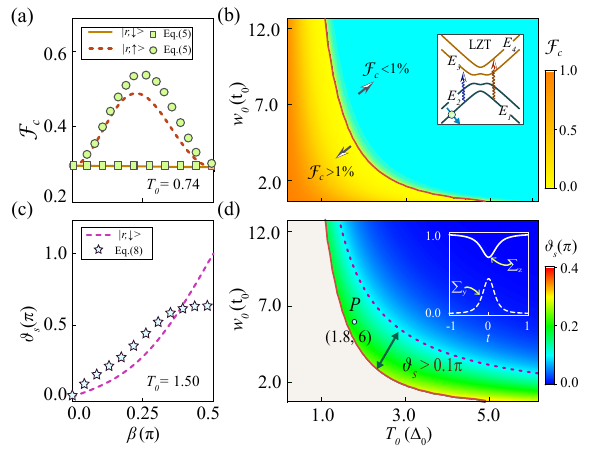}\\
  \caption{(a)   The infidelity of an inter-dot transfer  ${\cal F}^{}_{ c}$ as a function of     $\beta^{}_{  }$ for a qubit initialized in different spin states  of the  right QD, with   $V^{}_{0}=10$, $T^{}_{0}=0.74$, and $w^{}_{0}=7.2  t^{}_{0}$.  In addition to the numerical results (see lines), the discrete   symbols specify the   approximations from Eq.~(\ref{InF}).  (b)   The magnitude of  ${\cal F}^{}_{c}$ in terms of  $T^{}_{0}$ and $w^{}_{0}$ for an initial  spin-down  qubit with $\beta^{}_{ }=0.2\pi$. The yellow- and cyan-colored regions correspond to the parameter ranges with ${\cal F}^{}_{c} >1\%$ and ${\cal F}^{}_{c}<1\%$, respectively, while the solid curve denotes the boundary between these two regions.    Besides, the inset exhibits  the  possible LZTs.     (c) For an initial spin-down qubit,  the value of its spin-deviation angle $\vartheta^{}_{s}$   as a function of  $\beta$, with  $T^{}_{0}=1.5$ and    $w^{}_{0}=7.2t^{}_{0}$. Here, the dashed curve represents the numerical results, while the discrete symbols  specify the analytic results from Eq.~(\ref{lms}).    (d)     The numerical distribution of   $\vartheta^{}_{s}$   within the  range of ${\cal F}^{}_{c}<1\%$ in (b), with the doubled-head arrow signifying  the region of $\vartheta^{}_{s}>0.1\pi$.  In addition,     the    inset  exhibits  the  time dependence   of   $\Sigma^{}_{y,z}$   at the $P$ point  with $T^{}_{0}= 1.8$ and $w^{}_{0}=6t_{0}$. }\label{Fig3}
\end{figure}

   Using the fact of   $|\Psi^{}_{3/4} (w^{}_{0})\rangle \doteq |r,\downarrow/\uparrow\rangle$,  the transitions to the  higher-energy  $E^{}_{3,4}$   levels actually  prevent a leftward transfer, named as the  LZTs.  Quantitatively,  the  infidelity of the inter-dot  transfer  can be estimated as  ${\cal F}^{}_{c} =  |C^{}_{3}(w^{}_{0})|^{2}_{}+ | C^{}_{4}(w^{}_{0})|^{2}_{}$. As compared with   the  LZTs for a charged   qubit~\cite{Cao2013,Ota2018},    the presence  of    spin-dependent    tunnelings  between the QDs complicates the  transition processes. Dependent on the initial spin state of the qubit,      Fig.~\ref{Fig3}(a) shows that ${\cal F}^{}_{c}$ exhibits a  distinct    feature  versus $\beta$. It actually
       depict  the difference in the QDs'  spin-quantization axes, i.e., $\beta= \theta^{}_{\delta}$,   for $\varphi_{\rm so}=0$.  Specifically, ${\cal F}^{}_{c}$ is shown to be insensitive to the change of      $ \beta$    when the qubit   is  preset in   the  spin-down  state $|r,\downarrow\rangle$,  i.e., loaded  in the   $E_1$ energy level.  By contrast,   ${\cal F}^{}_{c}$  displays a strong  dependence of $ \beta^{}_{}$   if  prepared  in the spin-up state    $|r,\uparrow\rangle$, i.e., set in the  $E^{}_{2}$   level.  Clearly,  ${\cal F}^{}_{c}$ can attain  its peak  around $\beta^{}_{ }=\pi/4$   under  the  coordination  of the ``spin-flipped''   and  ``spin-conserved''  tunnelings.

By analogy to the  conventional  LZT theory~\cite{Wittig2005,Shevchenco2010,Sun2015,Krzywda2025}, the  infidelity  of an inter-dot transfer is dependent on the  rate of change of   $\epsilon$,  i.e.,  $\eta^{}_{0} =\partial \epsilon (t) /\partial t$  at $t=0$.   Thereby, the respective dependence relations in Fig.~\ref{Fig3}(a)  can be approximately evaluated as% approximately
 \begin{align}
 {\cal F}^{}_{c} \simeq  \begin{cases}
 e^{-\pi T^{2}_{0}/(\hbar \eta^{}_{0} ) }_{} ~~~~~\forall ~~~~~~~~ \Phi
  (-w^{ }_{0 }) =|r,\downarrow \rangle  \\
  \sum^{2}_{j=1} \gamma^{}_{j}e^{-\pi  \gamma^{}_{j}T^{2}_{0}/(\hbar \eta^{}_{0} ) }_{}   ~\forall ~  \Phi
  (-w^{ }_{0 })   =|r,\uparrow \rangle\ ,
 \end{cases}\label{InF}
 \end{align}
   with      $\gamma^{}_{1/2} = [1\pm \cos (2\beta )]/2$. Accordingly, the increase of $T^{}_{0}$ can  lead to the reduction of  ${\cal F} ^{}_{c}$.  Moreover, Fig.~\ref{Fig3}(b) shows the magnitude of  ${\cal F}^{}_{c}$ as a function of  $w^{}_{0}$ and $T^{}_{0}$.  Herein,  given that $\eta^{}_{0}=\pi V^{}_{0} / w^{}_{0}$,     the lower threshold of $T^{}_{0}$ for realizing  high-fidelity transfer    will  decrease with  the  extension of  $w^{}_{0}$. %,  as seen in  Fig.~\ref{Fig3}(b).

Upon achieving high-fidelity transfer,  it is important to note that in Fig.~\ref{Fig2}(d),    both $|C^{}_{1 } |^{2}_{}$ and  $|C^{}_{2 }  |^{2}_{}$  remain substantially non-zero in the final stable state ($t=w^{}_{0}$). This observation directly indicates a deviation of the transferred qubit  from  the  spin quantization axis  of the target (left) QD.   Addressed concretely,  based on the identity $|\Psi^{}_{1/2} (w^{}_0)\rangle \doteq |l,  \downarrow/\uparrow\rangle $,    the  deviation angle     can be calculated as
\begin{align}
 \vartheta^{}_{s}= 2\arccos\left(|C^{}_{2}(w_{0})/C^{}_{1}(w^{}_0) |\right)\ .
\end{align}
 In addition, we can  calibrate  the azimuthal angle of the transferred qubit, defined  within the  Bloch sphere  of the target QD  [see    Fig.~\ref{Fig2}(b)],  using the relation $\varphi^{}_{l} =2\phi^{}_{0} + \varphi^{}_{z}$ with $\varphi^{}_{z}=  \arg[C^{}_{2}(w^{}_{0})/C^{}_{1}(w^{}_{0})]$.
In fact,  deviation  from the local    spin quantization axis ($\hat{n}^{}_{l}$)  stems from  the interplay of  spin-dependent  inter-dot tunnelings.    Figure~\ref{Fig3}(c) illustrates  that  the  spin orientation  of a transferred  qubit    tends to be parallel ($\vartheta^{}_{s }=0$) or anti-parallel ($\vartheta^{}_{s }=\pi$)  to the  local axis in the  absence of  the    ``spin-flipped''  or  ``spin-conserved'' tunneling , i.e.,  when $\beta^{}_{ }=0$  or $\pi/2 $. From an alternative prospective,  it  indicates   two  parallel or antiparallel   spin quantization axes  of the   QDs for   $\varphi^{}_{\rm so}=0$.
 More interestingly,   Fig.~\ref{Fig3}(d) illustrates that  at  a fixed  (different) value of $\beta$,    the  deviation  angle $\vartheta^{}_{s}$     can  also   be regulated by the  two controllable parameters
       $T^{}_{0}$ and $w^{}_{0}$.
Herein, the  lower limit of  $ \vartheta^{}_{s}$  (where $ \vartheta^{}_{s} =0 $)  can be reached  as $T^{}_{0}$ or  $w^{}_{0}$ $\rightarrow \infty$, corresponding to the adiabatic limit.  The   upper limit of $\vartheta^{}_{s}  $ is  approximately set  by   the disparity between the local spin axes in Fig.~\ref{Fig1}(b), i.e., $  |\theta^{}_{l}-\theta^{}_{r} |$, as the parameters are tuned toward  the  threshold  boundary for    high-fidelity transfer.

%between the spin-quantization axes of the QDs,  i.e.,

%\sin(\gamma^{}_{1}+\gamma^{}_{2}) \dot{ \alpha}(t)
From an analytical standpoint, the    variation trend  of  $\vartheta^{}_{s }$     can be   captured by the dynamics of an effective    two-level   system (TLS), owing to  the suppression of  transitions to the $E^{}_{3,4}$ levels. Using Eq.~(\ref{dsf}), the effective  Hamiltonian describing the TLS defined in the   2D lower-energy state subspace   $\{|\Psi^{}_{1}(t)\rangle, |\Psi^{}_{2}(t)\rangle\}$ is given by
\begin{align}
H^{}_{\rm eff} =\Sigma^{}_{z} (t) \left(\begin{array}{cc}
   1&0\\
   0&-1\end{array}\right)+\Sigma^{}_{y} (t) \left(\begin{array}{cc}
   0&-i\\
   i&0\end{array}\right)\ ,\label{heff}
\end{align}
in which $\Sigma^{}_{ z}(t) = (E^{}_{1}   -E^{}_{2} )/2$ and  $\Sigma^{}_{y}(t) =\hbar {\cal T}^{}_{1,2}$ indicate the   longitudinal and transversal  components of an external driving  field.  Besides,
 the inset of  Fig.~\ref{Fig3}(d)  exhibits the specific  time evolutions of $\Sigma^{}_{y,z}(t)$    at $T^{}_{0}=1.8$ and $w^{}_{0}=6t^{}_{0}$, with $\beta=0.2\pi$.
By contrast to the persistence of
        $\Sigma^{}_{z}(t)$, it is seen  that  a considerable  $\Sigma^{}_{y}(t)$  can only exist  around  the zero time [see the inset of Fig.~\ref{Fig3}(d)]. Consequently,  the   TLS,  as preset in the ground state  ($|\Psi^{}_{1}\rangle$) at $t=-w^{}_{0}$,    will undergo a transient  oscillation.   %Based on the distinct major component   ($|l,\downarrow/\uparrow \rangle$) in $|\Psi^{}_{1/2}(w^{}_{0})\rangle $,
      In accordance with the major component ($ |l,\downarrow/\uparrow \rangle$)  in $|\Psi^{}_{1/2}(w^{}_{0})\rangle$,  the  mixing of  the two basis states   in its final state  can be   used  to quantify the
     spin orientation   of  a  transferred qubit.
In addition, for  $\beta\in[0,\pi/2)$  the spin-deviation angle  $\vartheta^{}_{s}$ can  be   approximated by the tilting   degree  of   the   driving  field   at    the zero time
     \begin{align}
     \vartheta^{}_{s} \simeq  \frac{2V^{}_{0}\xi^{}_{0}}{\sqrt{V^{2}_{0}\xi^{2}_{0} + 4 T^{2}_{0}\Sigma^{2}_{z}(0) w^{2}_{0}\cos^{2}_{}\beta^{}_{}}}\ ,\label{lms}
     \end{align}
     with   $\xi^{}_{0} = \hbar\pi (\sqrt{\iota^{}_{1}} +\sqrt{\iota^{}_{2}})/2 $  and $\iota^{}_{1/2}= [1\pm \cos (2\chi^{}_{1})] [1\mp \cos(2\chi^{}_{2})] $ at $\epsilon=0$. The details are provided in Appendix~\ref{S-2}.  It follows that   the  vanishing of $\xi^{}_{0}$ at  $\beta=0 $   prevents the spin deviation, and the     upward trend  of $\vartheta^{}_{s }$  with $\beta$   in Fig.~\ref{Fig3}(c)
    can be   (roughly)  explained by   the presence of $\cos^{ }_{}\beta$ in Eq.~(\ref{lms}).  Analogously, the difference  from the local quantization axis    is   also expected to   enlarge  with  the  decline   of $T^{}_{0}$  or/and $w^{}_{0}$.
 However, a continuous  reduction in these parameters will push the system beyond the viable operational  regime for realizing high-fidelity inter-dot transfer [see Fig.~\ref{Fig3}(b)], a prerequisite for studying  the  deviation of a transferred qubit from the local spin  axis.
Given this constraint, it imposes a fundamental limit on the attainable magnitude of  $\vartheta^{}_{s}$.  For a fixed value of $w^{}_{0}$ (or $T^{}_{0}$),     the  maximum attainable value of  $\vartheta^{}_{s}$   can then be achieved as  $T^{}_{0}$ (or  $w^{}_{0}$) approaches   the threshold boundary [see Fig.~\ref{Fig3}(d)].

%in terms of  the local spin  basis
\begin{figure}
  \centering
  \includegraphics[width=0.49\textwidth]{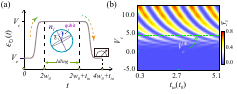}\\
   \caption{(a) The  time dependence of   $\epsilon^{}_{\circlearrowleft } (t)$ to implement one round of  inter-dot  transfer, in which  the up/down arrow    indicates the leftward/rightward shuttling  and  $t^{}_{in}$ is  the   time of a qubit idling in the left QD. In addition, both the Larmor precession  of qubit  stay in the left QD and the detection of its spin flipping  at  the  final moment are indicated. (b) The     spin-flipping probability  ${\cal F}^{}_{s}$  for  an initial spin-down qubit    as a function of $t^{}_{ in }$ and $V^{}_{c}$ with $\varphi^{}_{\rm so} =0.11\pi$,  $\theta^{}_{r}=0.44\pi$, $\theta^{}_{l}=0.95\pi$, $V^{}_{e} =-10$, $T^{}_{0}=2.0$, $w^{}_{0}=2t^{}_{0}$, and $V^{\star}_{c}$ being the threshold of the plateau value ($V^{}_{c}$)  for completing inter-dot   transfers.       }\label{Fig4}
\end{figure}

 Practically, $w^{}_{0}$ and $T^{}_{0}$ are the critical experimental parameters routinely tuned  to  control the spin state of a qubit  following inter-dot transfer~\cite{Wang2024,Doelman2024,Unseld2025}. Their specific values dictate whether the  transfer process approaches the non-adiabatic or adiabatic regime.   Here, by examining the evolution of qubit spin orientation across a range of parameter values, our study identifies   an intriguing optimal parameter regime favorable for the implementation of shuttling-based qubit gates.   This regime not only ensures high-fidelity transfer (i.e., ${\cal F}^{}_{c}<1\%$) but also guarantees a significant spin deviation (i.e., $\vartheta^{}_{s}>0.1\pi$), as highlighted by the double-headed arrow in Fig.~\ref{Fig3}(d). Evidently,  this finding also enriches the research on non-adiabatic spin dynamics of a qubit shuttled between    QDs with different Land\'{e} $g$-tensors.

Indeed, the spin states of a qubit before  and after a high-fidelity inter-dot shuttling  are interconnected by a transfer  matrix. More generally, the spin  state of   a qubit located in the left/right QD can be presented as  $|\psi^{}_{l/r} \rangle= \sum^{}_{\sigma=\uparrow,\downarrow }\nu^{}_{l/r, \sigma} |l/r,\sigma\rangle  $, with  $\boldsymbol{\nu}^{}_{ \kappa=l,r}=\{\nu^{}_{\kappa, \downarrow},\nu^{}_{\kappa,\uparrow}\}^{\rm T}_{}$  indicating the combination coefficients.  %For a fixed set of the control parameters    $w_{0}$ and $T^{}_{0}$,
  The effect of a leftward/rightward    shuttling   is reflected by the relation $ \boldsymbol{\nu}^{}_{l/r} =\hat{S}^{}_{l/r} \boldsymbol{\nu}^{}_{r/l}$, with $\hat{S}^{}_{\kappa=l,r}$ indicating the connection matrix.  Besides  the formation of  spin deviation from the local quantization axis,   it is important to notice that nonzero local Zeeman splitting  can render spin-up and spin-down qubits to accumulate different dynamic phases, i.e., $\pm\phi^{}_{z}$ with $\phi^{}_{z}\propto g^{}_{0}  w^{}_{0}   $ and $g^{}_{0}=\Delta^{}_{0}/\hbar$,  during the shuttling. Accordingly, $\hat{S}^{}_{\kappa}$ is shown to have the form of
\begin{align}
  \hat{S}^{}_{\kappa} =\left(\begin{array}{cc}
  e^{i \phi^{}_{\kappa}}_{}\cos \frac{\vartheta^{}_{s}}{2} &~- e^{-i\Theta^{}_{\kappa} }_{}\sin\frac{\vartheta^{}_{s}}{2}\\ \\
  e^{i \Theta^{}_{\kappa}}_{}\sin \frac{\vartheta^{}_{s}}{2}&~  e^{-i\phi^{}_{\kappa}}_{} \cos \frac{ \vartheta^{}_{s}}{2}
  \end{array}\right)\ , \label{Shum}
\end{align}
with $\phi^{}_{\kappa=l/r} = \phi^{}_{z} \pm  (\phi^{}_{\delta} -\phi^{}_{0})$, $\Theta^{}_{ \kappa} = \phi^{}_{\kappa}+\varphi^{}_{\kappa}$,   and $\varphi^{}_{ r}= 2\phi^{}_{\delta}+\varphi^{}_{z}+\pi$.  Thus far, we have established a basic framework to  fully capture the  spin dynamic  evolution  of a transferred qubit  undergoing  non-adiabatic inter-dot shuttling.    Building on this foundation, we are now poised to explore the practical implementation of shuttling-based spin qubit gates.

  \section{Implementation of shuttling-based  spin qubit gates}%Enabled by Larmor precession of a transferred qubit deviated from the local spin-quantization axis~\cite{Doelman2024}, shuttling-based spin qubit gates obviate high-frequency drivers, a key advantage over conventional AC modulation techniques.
 In contrast to  the unidirectional     transport scheme reported in Ref.~\onlinecite{Li2018a},  the shuttling-based single-qubit gates are implemented by returning the qubit to its initial QD. Evidently, this design can effectively mitigate gate infidelity induced by spatial inhomogeneities in system parameters.
    For a semiconductor circuit composed of two (or more) QDs, spin-qubit initialization and readout can be conveniently realized by leveraging the Pauli spin blockade effect~\cite{Wang2024,Petta2010}, which directly encodes the qubit spin state into a measurable charge signal.  Interestingly,   for an initial  spin-polarized qubit,  the deviation from the local quantization axis of one target QD can be manifested by  its spin-flipping signal   following a single round of inter-dot transfer~\cite{Doelman2024}.% returning  to the initial QD. %
   % probability of versus the magnetic

Addressed concretely,  the control of the time idling in the target  QD  is critical for determining the spin-flipping probability ${\cal F}^{}_{s}$.
 For brevity,   the   detuning required to implement one round of   anticlockwise  inter-dot  transfers is   modulated as  $\epsilon^{}_{\circlearrowleft} (t) = V^{}_{e}+ V^{}_{a}\{ f^{}_{1}(t) \tilde{H}(q^{}_{0}-t)+f^{}_{2}(t)[1-\tilde{H}(q^{}_{0}-t)]\}$, as  seen in Fig.~\ref{Fig4}(a). Here,   $f^{}_{1/2} (t) =(\tanh[\pi (q^{}_{1/2}\pm t)/w^{}_{0}]+1)/2  $, $q^{}_{0}=2w^{}_{0}+t^{}_{in}$, $q^{}_{1}=-w^{}_{0}$, $q^{}_{2}=3w^{}_{0}+t^{}_{in}$,     $\tilde{H}(t)$  is the Heaviside function, and   $t^{}_{in}$  represents the middle idling time. Correspondingly, Fig.~\ref{Fig4}(b) shows  the spin-flipping probability $ {\cal F}^{}_{s}$ for an initial spin-down qubit
 as a function of  $t^{}_{in}$ and  $ V^{}_{c}\equiv V^{}_{a}+V^{}_{e}$ at $t=t^{}_{in}+4w^{}_{0}$.  It is seen that   $ {\cal F}^{}_{s}$ can   oscillate with   $t^{}_{in }$  upon  the  completion of an inter-dot  transfer,
   as indicated by $V^{}_{c}> V^{\star}_{c} $ in Fig.~\ref{Fig4}(b). In this case, the effect of   leftward /rightward  unidirectional shuttling can be captured by  the respective connection matrix $\hat{S}^{}_{l/r}$  in Eq.~(\ref{Shum}).
 Meanwhile, due to the presence of a nonzero local Zeeman splitting,    a  short stay  in  one target  QD of  $t$ time  can be mapped to  an Z-gate operation   $\hat{Z}^{}_{}(t^{}_{ }) = {\rm diag} \{e^{ig^{}_{0} t^{}_{ } }_{} ,e^{-ig^{}_{0} t^{}_{  } }_{}\} $~\cite{Qi2024,Veldhorst2015,Russ2018,Wu2002}. Then,   the    operation matrix  for describing the   one   round  of  inter-dot transfer   becomes  $
   \mathbf{U}^{}_{ \circlearrowleft} = \hat{S}^{}_{r} \hat{Z}^{}_{} (t^{}_{in}) \hat{S}^{}_{l}  $, by which
   the flipping probability    can be estimated as
   \begin{align}
       {\cal F}^{}_{s}  =  \sin^{2}_{}\vartheta^{}_{s} \cos^{2}_{} (\phi^{}_{d} +g^{}_{0} t^{}_{in})\ ,\label{fs}
       \end{align}
   with $\phi^{}_{d}= \phi^{}_{r}+(\varphi^{}_{r}  -\varphi^{}_{l})/2$.   Please refer to Appendix~\ref{S-3} for the detailed derivations.
With the extension of $t^{}_{in}$ ,    it follows that  the oscillation amplitude of ${\cal F}^{}_{s}$ can be  used to quantify   the deviation angle  $\vartheta^{}_{s}$. Specifically,   the  raising of the  amplitude    with the increase of   $ V^{}_{c}$    in Fig.~\ref{Fig4}(b)    can   be  principally illustrated by the positive  dependence
 of  $\vartheta^{}_{s}$ on  $V^{}_{0}$    in Eq.~(\ref{lms}).      The  concurrent  tilting of the  oscillating stripes is  resulted from   the      variations in    $\phi^{ }_{d}$.    In addition,  the consistency with  the  detections in Fig.~2(b) of Ref.~\onlinecite{Doelman2024}   validates    our  theoretical   analyses.

\begin{figure}
  \centering
  \includegraphics[width=0.49\textwidth]{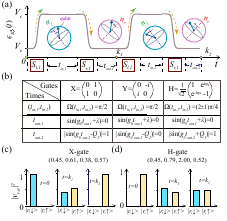}\\
  \caption{(a) The modulation sequence of $\epsilon^{}_{2\circlearrowleft}(t)$  to realize the  two rounds of inter-dot shuttling protocol,  in  which $t^{}_{in/out, n=1,2}$  indicates  the  time    idling     in the left/right QD     and   $\hat{S}^{}_{l/r,n}$  represents  the  connection matrix  of  the  leftward/rightward shuttling in the $n$-th round.   Therein,  $k^{}_{1}=4w^{}_{0}+t^{}_{in,1}+ t^{}_{out,1}  $, $k^{}_{2}=8w^{}_{0}+\sum^{2}_{n=1}( t^{}_{out,n} +t^{}_{in,n}) $, and the Larmor precessions of a qubit idling in the  QDs are also indicated. (b) Explicit equations for determining the  requisite idling times $t^{}_{in, 1}$, $t_{in,2}$,  $t^{}_{out,1} $ and $t^{}_{out,2}$   to implement  the single-qubit X, Y, and    generalized Hadamard (H) gates.     (c) and (d)  In the implementation of the  shuttling-based X and H gates,    the  composition ratios $|\nu^{}_{r,\sigma}|^{2}_{}$   of the local basis state  $|r,\sigma\rangle$ in the   evolving state of a  spin-polarized  qubit  at $t=0$ and $k^{}_{1,2}$.   Besides, the  specific    values of   $(t^{}_{ in,1}, t^{}_{out,1} ,t^{}_{in,2}, t^{}_{out,2} )$, in units of $t^{}_{0}$, are indicated in the panels and with the other parameters fixed as    $\varphi^{}_{\rm so} =0.11\pi$,  $\theta^{}_{r}=0.44\pi$, $\theta^{}_{l}=0.95\pi$, $T^{}_{1}=2.0$, $T^{}_{2}=2.2$, $ w^{}_{0}= 5t^{}_{0}$,   and $V^{}_{ c/e}=\pm10$.}
\label{Fig5}
\end{figure}

 Predictably,   the Pauli-X and Pauli-Y gates for a single spin qubit are implemented along the flipping of its initial spin orientation, a condition that corresponds to   ${\cal F}^{}_{s} =1$  in  Eq.~(\ref{fs}).    While for the general scenario of $\vartheta^{}_{s}\neq \pi/2$, the inequality ${\cal F}^{}_{s} <1$ precludes the realization of single-qubit gates by using the one-round inter-dot shuttling protocols. It should explore more rounds of inter-dot transfer processes to implement single-qubit gates in practical scenarios~\cite{Wang2024,Unseld2025}. For simplicity,  herein the gates are assumed to be realized based on two-round bidirectional shuttling protocols.
  Explicitly, the time evolution   of the desired detuning $\epsilon^{}_{2\circlearrowleft}(t)$ to implement the protocols is displayed in Fig.~\ref{Fig4}(a), in which       $t^{}_{in/out, n=1,2}$   denotes the time of the qubit idling in the left/right QD  in the $n$-th round.    The concrete  form of $\epsilon^{}_{2\circlearrowleft}(t)$
 can refer to Appendix~\ref{S-3}.
 Similar to Eq.~(\ref{Shum}),  we can use  $\hat{S}^{}_{l/r, n}$ to represent the connection matrix of  the leftward/rightward  inter-dot shuttling in the $n$-th round.  Meanwhile,  due to  the modulation in the corresponding tunnel-coupling  strength $T^{}_{n}$, the connection matrices  of the two rounds   are characterized by the distinct  $\vartheta^{}_{s,n}$, $\phi^{}_{l/r,n}$, and $\varphi^{}_{l/r,n}$, respectively.  Besides,   the specific  values  of these characteristic parameters are listed in Table~\ref{tab1}  for different  values of  $T^{}_{n }$ and $w^{}_{0}$. Expressed in the  local spin basis    $ \{|r,\downarrow\rangle,|r, \uparrow\rangle\}^{\rm T}_{}$ of the original QD,    the overall   operation matrix  describing  the effect of the  shuttling protocol is then constructed as
\begin{align}
\mathbf{U}^{}_{2\circlearrowleft}=&\hat{Z}^{}_{} (t^{}_{out,2}) \hat{S}^{}_{r,2} \hat{Z}^{}_{}(t_{in,2}) \hat{S}^{}_{l,2}
 \hat{Z}^{}_{} (t^{}_{out,1}) \nonumber\\ &\cdot\hat{S}^{}_{r,1} \hat{Z}^{}_{}(t^{}_{in,1}) \hat{S}^{}_{l,1}
\ .
\label{UST}
\end{align}

 In accordance with the characteristics  of the connection matrices $\hat{S}^{}_{l/r, n }$,  it is noted that a fine tuning of the  idling times $t^{}_{in,n}$ and  $t^{}_{out,n }$   is  crucial to the  implementation of    the   single-qubit gates,  as exhibited in Fig.~\ref{Fig5}(b).
  To facilitate the analysis,       two  $t^{}_{in,n}$-dependent  variables are   introduced as  $ A^{}_{ n=1,2 }   =  \cos \zeta^{}_{n}\cos\vartheta^{}_{s,n}  +i \sin \zeta^{}_{n} $, where $\zeta^{}_{n}= g^{}_{0} t^{}_{in,n} +\phi^{}_{d,n}$ and $\phi^{}_{d,n}=\phi^{}_{r,n } -(\varphi^{}_{l,n}-\varphi^{}_{r,n})/2$.  Then, the requirement for the   time    idling  in the right QD in the first round   can  be reflected by the formula
  $ \sin( g^{}_{0}t^{}_{out,1} +\lambda^{}_{}) =0$, with $\lambda^{}_{}= \eta^{}_{s} + \phi^{}_{l,2} -\varphi ^{ }_{ 12,c} $, $  \varphi ^{  }_{ nn^{\prime}_{},  c/s}=
(\varphi^{}_{r,n}\mp \varphi^{}_{l,n^{\prime}_{}})/2$, and  $  \eta^{}_{c/s}
=[\arg(  A^{}_{1})\mp\arg(  A^{}_{2})]/2 $. Under this circumstance,
      the operation matrix  in Eq.~(\ref{UST}) can  be further simplified into
 \begin{align}
\mathbf{U}^{}_{2\circlearrowleft}  =\left(\begin{array}{cc}
\cos \Omega e^{ig^{}_{0} t_{out,2}+iQ^{}_{1}}_{} &-\sin \Omega e^{ig^{}_{0} t_{out,2}-iQ^{}_{2}}_{}  \\
\sin \Omega e^{-ig^{}_{o} t^{}_{out,2}+iQ^{}_{2}}_{} & \cos\Omega e^{-ig^{}_{0} t_{out,2}-iQ^{}_{1}}_{}
\end{array}\right)\ ,\label{us2}
\end{align}
in which $\Omega (t^{}_{in,1},t^{}_{in,2}) =\sum^{}_{n }\arccos|  A^{}_{n} |$  and  $Q^{}_{1/2}=\eta^{}_{s/c}\mp \varphi^{}_{21,c/s}+\phi^{}_{l,1}$. It is evident that the single-qubit Pauli-X  gate  can   be attained if ($t^{}_{in,1}, t^{}_{in,2}$) satisfy the condition of $\Omega=\pi/2$ and  $|\sin( g^{}_{0}t^{}_{out,2}-Q_{2})|= 1 $.  Furthermore, the Pauli-Y gate   is obtained  by  adjusting  $t^{}_{out,2}$ to  meet the  corresponding  condition  $|\sin( g^{}_{0}t^{}_{out,2}-Q_{2})|=0$. Interestingly,    a generalized  Hadamard (H) gate  in Fig.~\ref{Fig5}(b) can  also  be realized  when $\Omega=  (2\pm1) \pi/4$    and   $|\sin(g^{}_{0}t^{}_{out,2}+Q^{}_{1})| =1$,  where   the  off-diagonal  phase shift  is  estimated as    $ \varphi^{}_{o}=  \tilde{\varphi}^{}_{}\mp \pi/2$  with  $\tilde{\varphi}^{}_{}=  \pi/2- Q^{}_{1}-Q^{}_{2}$. The  detailed derivations are presented  in  Appendix~\ref{S-3}.

 For  a qubit initialized in the spin-down   state  of the  right QD at $t=0$,
Fig.~\ref{Fig5}(c)    shows   the variations in the  composition ratio $|\nu^{}_{r,\uparrow/\downarrow}|^{2}_{}$ of the local basis state $|r,\uparrow/\downarrow\rangle$  in the qubit evolving state  at $t =k^{}_{1}$ and $t=k_{2}$. These two time points correspond to the completion of the first and second rounds of inter-dot transfer, respectively.
   Equipped with a  precise control of  the idling  times ($  t^{}_{in,1 }$ ,$t_{in,2}$,  $t^{}_{out,1} $,  $t_{out,2}$), it is seen that
the spin orientation of the   qubit   is  reversed, i.e., $\nu^{}_{r ,\downarrow }(k_2)=0$ and $|\nu^{}_{r ,\uparrow }(k_2)|^{2}_{}=1$, after going through the X-gate modulation.    In addition, Fig.~\ref{Fig5}(d) illustrates that  a generalized H gate can also  be attained with  tuning  $t^{}_{ out,1  }$, $t^{}_{in,1}$, and  $t_{out,2}$. For  a   qubit initialized  in the spin-up state of the right QD,  the composition ratios are found to be equal in the final state, i.e., $\nu^{}_{r ,\uparrow/\downarrow }(k^{}_{2})=1/2$,  and    a qubit superposition state of the form $ (e^{i\varphi^{}_{0}}_{}|r,\downarrow\rangle -  |r,\uparrow\rangle)/\sqrt{2}$ is  thus generated following the H-gate control sequence.

%

%%%%%%%%%%%%%%%%%%%%%%%%%%%%%%%%%%%%%%%%%%%%%%%%%
%%%%%%%%%%%%%%%%%%%%%%%%%%%%%%%%%%%%%%%%%%%%%%%%%

\section{Discussion and Conclusion} %\noindent{\color{blue}\bf{ Discussion and Conclusion---}}
  To date,    the effect of spin decoherence  have been neglected in the theoretical modeling of shuttling-based single-qubit gates.   Indeed,  this simplification is   well-founded  on the basis of a long  coherence time of the spin qubit.  For instance, the  coherence times  of  a hole spin qubit   in the Ge-based QDs in Ref.~\onlinecite{Wang2024}  have reached to a few of   microseconds. By contrast, the ramping  time of  the detuning  in experiments is on  the order of nanoseconds~\cite{Doelman2024,Wang2024,Unseld2025},  far less than   the  scale of the  coherence time. Thus, the spin qubit can maintain excellent coherence during multiple rounds of inter-dot transfer for realizing   qubit gates.
Especially, a significant deviation of the transferred qubit from the  spin-quantization axis of its residing QD is critical to the design of shuttling-based single-qubit gates.   For  the case    of   $\vartheta^{}_{s,n=1,2 }<\pi/4$, the inequality  $\Omega(t_{in,1},t_{in,2}) \neq\pi/2$ excludes the implementation of  the      X and Y  gates using the two  rounds  of inter-dot shuttling  protocol.
In such cases, it  should  explore    more rounds of  inter-dot transfer to realizing the anticipated logic gates, but a short  and brief   modulation  protocol    is  much  preferable to experiments.

 Accordingly, in this work, we thoroughly study  the change in the spin  orientation of a qubit after undergoing a  non-adiabatic shuttling between two QDs with different spin quantization axes. With the  realization of   a high-fidelity inter-dot transfer,   we  reveal that
  the deviation  from the local spin quantization axis is controllable, and   can be effectively regulated by   the tunnel-coupling strength  and the ramping time  of the detuning.
In addition,  based on the characteristics of    two rounds of the  designed  inter-dot
shuttling, we estimate  the requisite idling times   for realizing the   single-qubit  X,  Y and  generalized Hadamard  gates.
   Overall,   we  believe our  study will  invoke  more interest regarding  the non-adiabatic control of the spin orientation of a   transferred qubit and  the implementation of  high-fidelity   shuttling-based quantum logic gates in  semiconductor QD arrays.

%%%%%%%%%%%%%%%%%%%%%%%%%%%%%%%%%%%%%%%%%%%%%%%%%
%%%%%%%%%%%%%%%%%%%%%%%%%%%%%%%%%%%%%%%%%%%%%%%%%
% based on the characteristics of the unidirectional  shuttling,---the control of  spin orientation of a shuttled qubit and

 \noindent{\color{blue}\bf{Acknowledgements---}}  This work is supported by the National
Natural Science Foundation of China (Grants No.  92565304
and No. 92165208). X.-F. Liu acknowledges support  from the National Natural Science Foundation of China (Grant No. 62501059 ) and  the National Key Research and Development Program of China (Grant No. 2025YFE0217400).

\appendix
\begin{widetext}
\section{The derivation of the spin-dependent tunneling  amplitudes    ~\label{S-1}}

%\begin{widetext} In accordance with the model illustrated in Fig.~,
%,    $\omega^{}_{0}$   the  frequency  of the harmonic   confinement,
In this Appendix, we outline the calculation procedures  for  determining the  spin-dependent amplitudes for  a   spin qubit    tunneling   between   two   QDs  with  different spin-quantization axes.  In addition, the constrains on   the magnetic-field magnitude  $ B$ for defining well-behaved spin qubits in QDs and the analytic forms of the higher-energy states  $|\Psi^{}_{3,4}\rangle$ are  presented.

%The general state of the spin-quantization  ---
In the presence of an external magnetic field    applied in the $y$ direction [see Fig.~\ref{Fig1}(a)],  the distinct spin states of a  single hole (or electron) confined in the QDs   undergo splitting due to the Zeeman effect.  Before proceeding further,the introduction of well-defined spin qubits in QDs actually imposes strict constraints on the field magnitude  $B$. To begin with, the magnetic-field  induced Zeeman splitting $\Delta^{}_{z}$ must be smaller than the  QD orbital energy-level splitting $\Delta^{}_{ }E^{}_{orb}$,   a fundamental condition for ensuring the encoding of pure spin qubits.  For a qubit of $\Delta E^{}_{orb}  \sim 1.2$meV; with $g\simeq0.4$ for Ge holes~\cite{Hendrickx2024}, the corresponding upper threshold for $B$ is 51~T  [calculated via   $B=\Delta E^{}_{orb}/(g\mu^{}_{\rm B})]$.   In addition, accounting for  the hyperfine interaction   with  nuclear spins,  the magnitude $B$  should be much larger than  the strength  ($B^{}_{hf}$) of the average hyperfine fields for mitigating its adverse impacts~\cite{Hung2013,Fischer2010,Testelin2009}.  Furthermore, when  considering the inverse-$B$ dependence of spin qubit relaxation times, the preservation of long qubit coherence times  also impose    an upper limit on the  value  of $B$~\cite{Wang2021,Sarkar2023,Golovach2004,Maier2013}.     For instance,    the  coherence times  of  Ge-based  hole spin qubits in experiments can   reach to the microsecond regime  as   $B\leq40$~mT  (while exceeding $B^{}_{hf}\sim  20\mu$T)~\cite{Wang2024}, and
 this prolonged coherence window effectively suppresses the formation of relaxation-induced unintended spin polarization in   subsequent qubit manipulations.

With including both the spin and spatial degrees of freedom,   the low-energy Zeeman splitting  states of  the   qubit  in the QDs can be formulated as
  \begin{align}
 |l/r,\uparrow\rangle= \psi^{}_{0}( x \pm d^{}_{} ,y ) |  s^{ + }_{l/r }\rangle ~~~ |l/r,\downarrow\rangle=  \psi^{}_{0} ( x \pm  d^{}_{} ,y) |s^{ -}_{l/r }\rangle\ .\label{was}
  \end{align}   Here,  $\psi^{}_{0} (x\pm d,y)$ indicates  the ground-state  wavefunction  of the left/right QD in the  $x-y$ plane, $2d$ quantities the   inter-dot distance along the $x$ direction, and  $|s^{\pm }_{l/r  }\rangle$  characterize the   spin  components. Because of the inhomogeneity in the  Land\'{e} $g$-tensor,  the orientation of their spin quantization axis  is different between  the left and right QDs, just as shown in Fig.~\ref{Fig1s}(a). More specifically, the local spin components in Eq.~(\ref{was}) satisfy the relation $(\hat{n}^{}_{\kappa}\cdot \boldsymbol{\sigma}) |s^{\pm}_{\kappa}\rangle =  \pm |s^{\pm}_{\kappa} \rangle$  and ,
        expanded in the  unified spin     basis  of  $ \{|\uparrow_{ }\rangle , |\downarrow_{ }\rangle\}^{\rm T}_{}$   with  $\sigma^{}_{z}|\uparrow_{}/\downarrow_{}\rangle =\pm |\uparrow_{}/\downarrow_{}\rangle$,   they can be written as
  \begin{align}
  |s^{+}_{\kappa }\rangle =  \frac{1}{\sqrt{2}}\left(\begin{array}{c}
  \cos \frac{\theta^{}_{\kappa}}{2} -i\sin\frac{\theta^{}_{\kappa}}{2}\\
   \cos \frac{\theta^{}_{\kappa}}{2} +i\sin\frac{\theta^{}_{\kappa}}{2}
  \end{array}\right) ~~|s^{-}_{\kappa }\rangle  =\frac{1}{ \sqrt{2}} \left(\begin{array}{c} i\sin\frac{\theta^{}_{\kappa}}{2}-\cos\frac{\theta^{}_{\kappa}}{2}\\
   \cos \frac{\theta^{}_{\kappa}}{2} +i\sin\frac{\theta^{}_{\kappa}}{2}
  \end{array}\right)\ . \label{swav}
  \end{align}

\begin{figure}
 \centering
%  Requires \usepackage{graphicx}
   \includegraphics[width=0.88\textwidth]{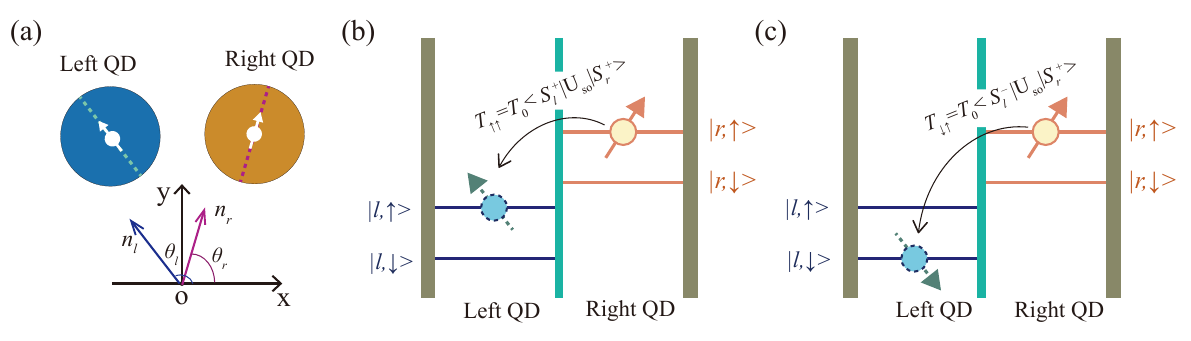}\\
   \caption{(a)  Schematic  of  the  spin-quantization  axes  of the left and right QDs,  i.e.,    $\hat{n}^{}_{l}= ( \cos\theta^{}_{l} ,\sin\theta^{}_{l},0)$  and $\hat{n}^{}_{r} =(\cos\theta^{}_{r} ,\sin\theta^{}_{r},0)$,  in the $x-y$ plane. (b)   The ``spin-conserved''  tunneling  of a   qubit    between the upper Zeeman splitting states $|l,\uparrow\rangle $ and $|r,\uparrow\rangle$ of the two QDs, with $T^{}_{\uparrow,\uparrow}$ indicating the tunneling amplitude. (c)  The ``spin-flipped''  tunneling for a   qubit transferring from  the  spin-up state  $|r,\uparrow\rangle$  of the right QD to the spin-down state  $|l,\downarrow\rangle$  of the other QD and with $T^{}_{\downarrow\uparrow}$ denoting the    amplitude.
   }\label{Fig1s}
 \end{figure}

 Theoretically, the strength of   tunnel-coupling    between  two  coupled  QDs, denoted by $T^{}_{0}$, is correlated to the overlap between   the  qubit's spatial   wave-functions $\psi^{}_{0} (x\pm d,y)$~\cite{Liu2018x}.  In experiments, this    can   be regulated by  the  middle barrier-gate potential~\cite{Tidjani2023}, see $V^{}_{M}$ in Fig.~1(a).  Given the difference in the spin-quantization axes of the coupled QDs,   the  spin-dependent    tunneling amplitude  $T^{}_{\sigma^{\prime}_{} \sigma}$   depends not only on  the tunnel-coupling strength, but also on  the disparity in the  spin  axes of the two QDs. Beyond these factors, it can also be regulated by the  SOI in the semiconductors. Specifically, the  Hamiltonian for the (linear) Rashba SOI along  the tunneling ($x$-axis) direction  takes the form of $H^{}_{\rm so}=\alpha^{}_{\rm so} p^{}_{x}\sigma_{y}$~\cite{Winkler2023}, where $p^{}_{x}=-i\hbar \partial /(\partial x)$ and $\alpha_{\rm so} $ denotes the strength of the SOI. In conjunction with inter-dot tunneling, the SOI effect behaves as a unitary operator acting on the transferred spin qubit~\cite{Liu2022}
    \begin{align} \mathbf{U}^{}_{\rm so} = \exp (i \varphi^{}_{\rm so}\sigma^{}_{y}   ) =\left(\begin{array}{cc}
  \cos\varphi^{}_{\rm so}~ &~  \sin\varphi^{}_{\rm so}\\
  -\sin\varphi^{}_{\rm so}~&
  ~\cos\varphi^{}_{\rm so}
  \end{array}\right)\ .
  \label{UOP}
   \end{align}Here,   $ \varphi^{}_{\rm so} = 2d/\ell^{}_{\rm so}$  indicates the  Aharonov-Casher (AC) phase,  $\ell^{}_{\rm so}=\hbar/(m^{}_{e}\alpha_{\rm so})  $  is  the  effective spin-orbit length, and $m^{}_{e}$ the effective mass of the charged particle.  Then,   the amplitude  for  a  qubit tunneling from    the  upper/lower  Zeeman-splitting  state of the right QD to the     upper/lower splitting  state of the left QD can be evaluated as $T^{}_{\uparrow\uparrow/\downarrow\downarrow} =T^{} _{0} \langle s^{\pm}_{l} |\mathbf{U}^{}_{\rm so}| s^{\pm}_{r}\rangle$, see Fig.~\ref{Fig1s}(b).    Based on  Eqs.~(\ref{swav}) and (\ref{UOP}), it can be further simplified into
  \begin{align}
 T^{}_{\uparrow\uparrow/\downarrow\downarrow}   = T^{}_{0}  \left( \cos\varphi^{}_{\rm so} \cos\frac{\theta^{}_{r}-\theta^{}_{l}}{2} \pm i\sin\varphi^{}_{\rm so}\sin \frac{\theta^{}_{r}+\theta^{}_{l}}{2} \right)\ .
 \label{TC}
  \end{align}
  with $\theta^{}_{\delta}=(\theta^{}_{r}-\theta^{}_{l})/2$ and $\theta^{}_{0}=(\theta^{}_{r}+\theta^{}_{l})/2$.
Analogously,  the  amplitude  for a  qubit  tunneling   from    the    lower/upper Zeeman-splitting  state of the right QD to the  upper/lower   splitting  state of the other QD is give by  $ T^{}_{\uparrow\downarrow/\downarrow\uparrow} = T^{} _{0} \langle s^{\pm}_{l} |\mathbf{U}^{}_{\rm so}| s^{\mp}_{r}\rangle$, see Fig.~\ref{Fig1s}(c).  According to the concrete expressions  of the spin components in  Eq.~(\ref{swav}),  these amplitudes  can be calculated as
  \begin{align}
 T^{}_{\uparrow\downarrow/\downarrow\uparrow}   = T^{}_{0}  \left( i\cos\varphi^{}_{\rm so} \sin\frac{\theta^{}_{r}-\theta^{}_{l}}{2}\pm  \sin\varphi^{}_{\rm so}\cos \frac{ \theta^{}_{r}+\theta^{}_{l}}{2} \right)\ .
 \label{TF}
  \end{align}

 By including  the  influence of inter-dot  detuning ($\epsilon$),   the  effective  Hamiltonian describing  the  single  qubit  in the double QDs  is given in Eq.~(\ref{H0}). To facilitate  the analysis, the spin-dependent tunneling amplitudes in Eqs.~(\ref{TC}) and (\ref{TF}) are    reformulated as
 \begin{align}
 T^{}_{\uparrow\uparrow  } =&T^{}_{0} \cos\beta e^{i \tilde{\zeta}^{}_{1}}_{} ~~ ~~T^{}_{\downarrow\downarrow} =T^{}_{0} \cos\beta e^{-i \tilde{\zeta}^{}_{1}}_{} \nonumber\\ T^{}_{\uparrow\downarrow} =&T^{}_{0} \sin\beta e^{i\tilde{\zeta}^{}_{2}}_{}~~~~T_{\downarrow\uparrow}^{} =-T^{}_{0}\sin\beta e^{-i\tilde{\zeta}^{}_{2}}_{}\ ,
  \end{align}
  with $\beta=\arccos(|T^{}_{\uparrow\uparrow}|/ T^{}_{0})$, $ \tilde{\zeta}^{}_{1}=\arg (T^{}_{\uparrow\uparrow})$, and $\tilde{\zeta}^{}_{2}=\arg(T^{}_{\uparrow \downarrow})$.  Based on this, a new normalized local    basis $\boldsymbol{\Pi}= \{ e^{i\phi^{ }_{0}}_{}|l,\uparrow\rangle, e^{-i\phi^{}_{0}}_{}|l,\downarrow\rangle, e^{i\phi^{ }_{\delta}}_{}|r,\uparrow\rangle, e^{-i\phi^{ }_{\delta}}_{}|r,\downarrow\rangle \}^{\rm T}_{}$   is introduced, with $\phi^{}_{0/\delta} =(\tilde{\zeta}_{2} \pm \tilde{\zeta}^{}_{1})/2$.     By projecting into the  new   basis, the  Hamiltonian in Eq.~(\ref{H0}) can be  expanded as
  \begin{align}
  H^{}_{0}(\epsilon) =  \left(\begin{array}{cccc}
  E^{}_{l\uparrow} & 0& T^{}_{0} \cos\beta & T^{}_{0}\sin\beta\\
  0& E^{}_{l\downarrow} &-T^{}_{0}\sin\beta &T^{}_{0}\cos\beta\\
 T^{}_{0}\cos\beta &-T^{}_{0}\sin\beta & E^{}_{r\uparrow} &0\\
 T^{}_{0} \sin\beta & T^{}_{0}\cos\beta &0 &E^{}_{r\downarrow}
 \end{array}\right)\ .
 \label{HM0}
  \end{align}
  Then,
 the     energies  $E^{}_{j=1-4} $  of the single qubit can be analytically derived through the direct diagonalization of Eq.~(\ref{HM0}), i.e.,
\begin{align}
E^{}_{1}  (\epsilon)=
- \sqrt{(\Delta^{}_{0}+\Gamma^{}_{0})^{2}_{}+T^{2}_{0}\sin^{2}_{}\beta}~~~~
E^{}_{2}(\epsilon) =- \sqrt{(\Delta^{}_{0}-\Gamma^{}_{0})^{2}_{}+T^{2}_{0}\sin^{2}_{}\beta}\nonumber\\
E^{}_{3}(\epsilon) = \sqrt{(\Delta^{}_{0}-\Gamma^{}_{0})^{2}_{}+T^{2}_{0}\sin^{2}_{}\beta}~~~~
E^{}_{4}  (\epsilon)=
  \sqrt{(\Delta^{}_{0}+\Gamma^{}_{0})^{2}_{}+T^{2}_{0}\sin^{2}_{}\beta}~~~
\label{enes}\ ,
\end{align}
with $\Delta^{}_{0}= \Delta^{}_{z } /2$   and $\Gamma^{}_{0}
=\sqrt{\epsilon^{2}_{}+T^{2}_{0}\cos^{2}_{}\beta}$. In accordance with  the  corresponding   eigenvectors  , the lowest two energy states $|\Psi^{}_{1,2}\rangle$ are given in Eq.~(\ref{psl}) of the main text.   Similarly, the  specific forms of the two higher-energy states  $|\Psi^{}_{3,4}(\epsilon)\rangle$ are presented by
 \begin{align}
|\Psi^{}_{3}(\epsilon)\rangle=\left(\begin{array}{c}
 \cos\alpha \cos\chi^{}_{2}\\
  \sin\alpha\sin \chi^{}_{2}\\
 - \sin\alpha\cos\chi^{}_{2}\\
  \cos\alpha \sin\chi^{}_{2}
  \end{array}\right)~~|\Psi^{}_{4}(\epsilon)\rangle =  \left(\begin{array}{c}
  \sin\alpha \sin\chi^{}_{1}\\
  - \cos\alpha \cos\chi^{}_{1} \\
   \cos\alpha \sin\chi^{}_{1}\\
  \sin\alpha \cos\chi^{}_{1}
  \end{array}\right)\ ,
  \label{cps}
 \end{align}
with  the specific expressions of $ \alpha^{}_{} $ and  $ \chi^{}_{n=1,2 }$ illustrated in the main text below Eq.~(\ref{psl}).

%Equations (\ref{tun1}) and
%scenario

%\begin{align}

%electron

\section{ The analytic expressions   of  the matrix elements  ${\cal T}^{}_{j,j^{\prime}_{}}(t)$    ~\label{S-2} }

In this Appendix , we outline  the  specific  expressions of  the matrix elements ${\cal T}^{}_{j,j^{\prime}_{}}(t)$  in  Eq.~(\ref{dsf}).  Besides, the derivation of the analytical approximation for the spin derivation  angle   in Eq.~(\ref{lms})   is   presented.

\begin{figure}
  \centering
  \includegraphics[width=0.58\textwidth]{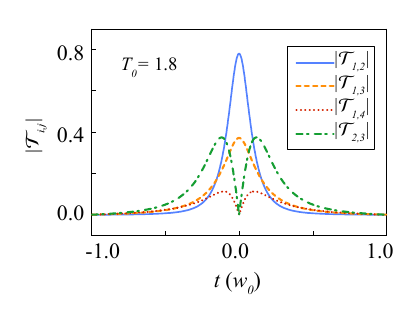}\\
  \caption{(a)  The  magnitudes of  the  transition  elements $|{\cal T}^{}_{1,2}|$, $|{\cal T}_{1,3}|$, $|{\cal T}^{}_{1,4}|$, and $|{\cal T}^{}_{2,3}|$ ( in   units  of  $ \pi/t^{}_{0}$ ) as  a function of  $t$  with  $T^{}_{0}= 1.8$, $V^{}_{0}=10$, $\beta=0.32\pi$, and  $w^{}_{0}=6t^{}_{0}$.
   }\label{Fig2s}
\end{figure}

 % as we focus on
 %%being different from the case
According  to  the orthogonality of the energy states  $|\Psi^{}_{j=1-4}(\epsilon)\rangle$,     the time evolution state of the qubit can be expanded  as    $ \Phi (t) =\sum^{4}_{j=1}C ^{}_{j}(t)|\Psi^{}_{j} (\epsilon)\rangle  $. Substituting it into  the time-dependent Schr\"{o}dinger equation
  \begin{align}
   i\hbar \frac{\partial }{\partial t} \Phi(t) = H^{}_{0} (\epsilon) \Phi(t) \ ,
    \end{align} it is found that the   combinational coefficients   $C^{}_{j}(t)$   are determined by the  the  differential equation (\ref{dsf}) of the main text
\begin{align}
i\frac{d C^{}_{j}(t)}{dt}+i\sum^{4}_{j^{\prime}_{}=1}{\cal T}^{}_{j, j^{\prime}_{}}(t)C^{}_{j^{\prime}_{}}(t) =\frac{E^{}_{j}(\epsilon )}{\hbar}C^{}_{j}(t) \ ,
\label{kjj}
\end{align}
with   ${\cal  T}^{}_{j,j^{\prime}_{}} (t) =  \langle \Psi^{}_{j} (\epsilon)| \partial ^{}_{t}| \Psi^{}_{j^{\prime}_{}} (\epsilon)\rangle $.  Based on  the concrete form of    $|\Psi^{}_{j}\rangle$ in Eqs. (\ref{psl}) and (\ref{cps}),  it is  found  that ${\cal T}^{}_{j,j^{\prime}_{}}(t)=-{\cal T}^{}_{j^{\prime}_{},j}(t)$ as such the off-diagonal terms ${\cal T}^{}_{j,j}=0$. The off-diagonal terms are derived by
\begin{align}
&{\cal T}^{}_{1,2} (t)  = -{\cal T}^{}_{2,1} (t)= \sin ( \chi^{} _{1}+\chi^{}_{2}) \frac{\partial \alpha}{\partial t}\nonumber\\  &{\cal T}^{}_{1,3}(t)=- {\cal T}^{}_{3,1}(t)= -\cos(\chi^{}_{1} +\chi^{}_{2})  \frac{\partial \alpha}{\partial t} \nonumber\\ &{\cal T}^{}_{1,4} (t)= -{\cal T}^{}_{4,1} (t)=\frac{\partial\chi^{}_{1}}{\partial t}\nonumber\\ &
{\cal T}^{}_{2,3} (t) = -{\cal T}^{}_{3,2} (t)=\frac{\partial \chi^{}_{2}}{\partial t}\nonumber\\ &{\cal T}^{}_{2,4} (t)= -{\cal T}^{}_{4,2} (t)=\cos (\chi^{}_{1}+\chi^{}_{2}) \frac{\partial\alpha}{\partial t}\nonumber\\ &{\cal T}^{}_{3,4} (t)= -{\cal T}^{}_{4,3} (t)=\sin (\chi^{}_{1}+\chi^{}_{2}) \frac{\partial \alpha}{\partial t}\ .
 \label{fms}
\end{align}
  Based on  the  time-dependent modulation  of the detuning  $\epsilon(t) = V^{}_{0}\tanh\left(\pi t/w^{}_{0}\right)$, and using the definitions of the   specific  parameters   $\alpha$ and $\chi^{}_{ 1,2}$,   one  can be  obtain that
 \begin{align}
\frac{\partial \alpha} {\partial t} = & -\frac{\pi}{2} \frac{T^{}_{0}V^{}_{0}\cos\beta}
 { \Gamma^{2}_{0}w^{}_{0}}{\rm sech}^{2}_{}(\frac{\pi t}{w^{}_{0}}) \nonumber\\ \frac{\partial\chi^{}_{n }}{\partial t} = & \frac{\pi}{2}\frac{V^{2}_{0}T^{}_{0}
 \sin\beta}{w^{}_{0}\Gamma^{}_{0}  E^{2}_{n}} {\rm sech}^{2}_{} (\frac{\pi t}{w^{}_{0}})
 \tanh(\frac{\pi t}{w^{}_{0}})\ .\label{alp}
 \end{align}
Accordingly, it is confirmed that  ${\cal T}^{}_{j^{\prime}_{},j^{}_{}} \propto {\rm sech}^{2}_{} (\frac{\pi t}{w^{}_{0}})$,  and   Fig.~\ref{Fig2s}  exhibits the specific  time evolutions of   $|{\cal T}^{}_{1,2}|$,  $|{\cal T}^{}_{1,3}|$,  $|{\cal T}^{}_{1 , 4}|$ and $|{\cal T}^{}_{2,3}|$. It can be observed that non-negligible nonzero values of
 $|{\cal T}^{}_{1,2}|$ and $|{\cal T}^{}_{1,3}|$   arise only near $t=0$, where
  $|{\cal T}^{}_{1 , 4}|$ and $|{\cal T}^{}_{2,3}|$  simultaneously drop to zero owing to the factor    $\tanh(\frac{\pi t}{w^{}_{0}})$ in the second line of  Eq.~(\ref{alp}) .

 %Uniqueness,   by exploiting the orthogonality of the energy states $|\Psi$ The

  % Based on

%%% how to present the  right answer with the different

Upon the realization of high-fidelity inter-dot transfer, it is demonstrated that the magnitudes of   $|C^{}_{3,4 }|^{2}_{} $ are significantly suppressed throughout the process, see Fig.~\ref{Fig2}(d) of the main text.   Building on this result and in accordance with Eq. (\ref{kjj}), the time evolutions of   $C^{}_{1,2} $   can be  captured  by a  simplified   matrix equation
 \begin{align}
&\frac{\partial}{\partial t} \left(\begin{array}{c}
C^{}_{1}(t)\\
C^{}_{2}(t)
 \end{array}\right) +\left(\begin{array}{cc}
0&{\cal T}^{}_{1,2} (t) \\
{\cal T}^{}_{2,1} (t) &0
\end{array}\right)\left(\begin{array}{c}
C^{}_{1}(t)\\
C^{}_{2}(t)
 \end{array}\right)=\frac{1}{i\hbar}\left(\begin{array}{c} E^{}_{1}(t)
C^{}_{1}(t)\\
E^{}_{2}(t)C^{}_{2}(t)
\end{array}\right)\ .\label{eds}
 \end{align}
 Intriguingly, Eq.~(\ref{eds})   is  found to be  equivalent to the  time-dependent Schr\"{o}dinger  equation for a two-level system (TLS) under an external driving field.  Working in the 2D lower-energy state subspace  $\{|\Psi^{}_{1}(\epsilon)\rangle ,|\Psi^{}_{2}(\epsilon)\rangle\}^{\rm T}_{}$, to be more specific, the  effective Hamiltonian   is given by Eq.~(\ref{heff}) of the main text, i.e.,
\begin{align}
  H^{}_{\rm eff} (t) =\Sigma^{}_{z}(t)\left(\begin{array}{cc}
  1&0\\
  0&-1
  \end{array}\right)+\Sigma^{}_{y}(y) \left(\begin{array}{cc}
  0&-i\\
  i&0
  \end{array}\right) ,\end{align} with$
  \Sigma^{}_{z}(t) = (E^{}_{1} -E^{}_{2} )/2 $ and $ \Sigma^{}_{y}(t)= \hbar {\cal T}^{}_{1,2}(t)
$
  indicating  the  longitudinal and  transverse  components of the driving  field. Using Eqs. (\ref{fms}) and (\ref{alp}),
      the transverse    component    can be  further evaluated as
   \begin{align}
     \Sigma^{}_{y} (t)= & -\frac{\pi}{4}\frac{ \hbar V^{}_{0}T^{}_{0} }{w^{}_{0}\Gamma^{2}_{0}(t)}  \cos\beta {\rm sech}^{2}_{}  (\frac{\pi t }{w^{}_{0}} )
      \Big\{\sqrt{  [1+ \cos(2\chi^{}_{1})][1- \cos(2\chi^{}_{2})] } \nonumber\\ &+\sqrt{ [1-\cos(2\chi^{}_{1})][1+ \cos(2\chi^{}_{2})] }\Big\}\ .
   \label{BBC}
   \end{align}

As shown in the inset of Fig.~\ref{Fig3}(d) in the main text,   $\Sigma^{}_{y}$  and $\Sigma^{}_{z} $  exhibit distinct time dependences.   In contrast to the longitudinal component $\Sigma^{}_{z}$,    it is seen that the presence of a non-negligible transverse component  $\Sigma^{}_{y} (t)$  is confined to the vicinity of
 $t=0$.  This time-dependent behavior accounts for the deviation of a leftward transferred qubit from the local spin quantization axis.
  For a  qubit  initialized in the spin-down state of the right QD, it is confirmed that the  deviation angle  can   be   approximated by the tilting degree of the field  at the central time, i.e., $\vartheta^{}_{s}\simeq  2\Sigma^{}_{y}(0)/ \sqrt{\Sigma^{2}_{z} (0) +\Sigma^{2}_{y}(0)}^{}_{}$. Using  the concrete form of $\Sigma^{}_{y}(t)$ in Eq.~(\ref{BBC}), it can be further simplified into Eq.~(\ref{lms}) of the main text
   \begin{align}
   \vartheta^{}_{s} \simeq \frac{2  V^{}_{0}\xi^{}_{0}}{\sqrt{  V^{2}_{0} \xi^{2}_{0}+4 T^{2}_{0} w^{2}_{0}
     \Sigma^{2}_{z}(0)\cos^{2}_{}\beta}}\ ,
   \end{align}
      in which  $\xi^{}_{0}=\hbar \pi (\sqrt{\iota^{}_{1}} +\sqrt{\iota^{}_{2}})/2$ and $  \iota ^{}_{1/2} $ denote the values of  $ [1\pm\cos(2\chi^{}_{1})][1\mp \cos(2\chi^{}_{2})] $  evaluated at $t=0$, respectively. In the case of $\beta=0$,    it is found that  $\vartheta^{}_{s}=0$  because   $ \iota^{}_{n=1,2} =0$.  For $\beta\neq 0$ (or $\pi/2$), the spin deviation angle $\vartheta^{}_{s}$ is dependent on the specific values of   $T^{}_{0}$ and $w^{}_{0}$,  as discussed in the main text.

\section{    The requisite idling times in realizing single-qubit gates ~\label{S-3}}

In this  Appendix, we present the details for ascertaining the required  idling times  in realizing the single-qubit X, Y, and generalized H gates. Before that,    we outline the calculation  process for determining    the spin-flipping probability ${\cal F}^{}_{s}$ of a spin qubit after undergoing one round of inter-dot transfer.% In addition,      implicit  equations      are evaluated  in the  implementation of the single-qubit  gates.

 Based on the explicit form of the connection matrix  $\hat{S}^{}_{l/r}$ given in Eq.~(\ref{Shum}) of the main text,   the operation matrix describing one round of inter-dot transfer---driven by the time-dependent detuning modulation $\epsilon$ shown in Fig.~\ref{Fig4}(a)---can be   expanded as
\begin{align}
   \mathbf{U}^{}_{ \circlearrowleft} = \hat{S}^{}_{r} \hat{Z}^{}_{} (t^{}_{in}) \hat{S}^{}_{l}=\left(\begin{array}{cc}
  e^{i\phi^{}_{l}}_{} {\rm M}^{}_{11}  & e^{-i\phi^{}_{l}}_{}{\rm M}^{}_{12} \\
  e^{i\phi^{}_{l}}_{}{\rm M}^{}_{21} & e^{-i\phi^{}_{l}} _{} {\rm M}^{}_{22}
   \end{array}\right)\ ,\label{mat}
   \end{align}
   in which  \begin{align}
  {\rm M}^{}_{11}= &e^{i \varphi^{}_{\delta} }_{} [\cos(
   g^{}_{0} t^{}_{in} +\phi^{}_{d})
   \cos\vartheta^{}_{s}+i\sin(g^{}_{0} t^{}_{in} +\phi^{}_{d})]\nonumber\\
    {\rm M}^{}_{12} =&-e^{-i\varphi^{}_{0}}_{} \sin\vartheta^{}_{s} \cos(\phi^{}_{d} +g^{}_{0}t^{}_{in})\ ,
    \end{align}
    ${\rm M}^{}_{22}={\rm M}^{\ast}_{11}$, and
   ${\rm M}^{}_{21} = -{\rm M}^{\ast}_{12}$, with $\varphi^{}_{0/\delta} =(\varphi^{}_{l}  \pm \varphi^{}_{r})/2$ and $\phi^{}_{d}= \phi^{}_{r}-\varphi^{}_{\delta}$.   Expressed in the local spin basis   $ \{|r,\downarrow\rangle, |r,\uparrow\rangle\}^{\rm T}_{}$, the   evolved  spin state of a qubit  initialized   in the  spin-down state  of the right QD  at $t=0$ can then  be calculated as
    \begin{align}
    |\psi^{}_{r}(t^{}_{f})\rangle=  \mathbf{U}^{}_{ \circlearrowleft} \left(\begin{array}{c}
    1\\
    0
    \end{array}\right)=  \left(\begin{array}{c}
      e^{i\phi^{}_{l}}_{} M^{}_{11} \\
   e^{i\phi^{}_{l}}_{}M^{}_{21}
    \end{array}\right)\ ,
    \end{align}
    with $t^{}_{f}=4w^{}_{0}+t^{}_{in}$.
  Consequently,  it follows that the spin-flipping probability   ${\cal F}^{}_{s}$ of  the transferred qubit can be estimated as
   \begin{align}
       {\cal F}^{}_{s} = |{\rm M}^{}_{21}|^{2} _{} =  \sin^{2}_{}\vartheta^{}_{s} \cos^{2}_{} (\phi^{}_{d} +g^{}_{0} t^{}_{in})\ .
       \end{align}

For the general case of $\vartheta^{}_{s} \neq 0$,  it is evident that the implementation of   the single-qubit Pauli-X  and Pauli-Y gates  necessitates multiple  rounds of inter-dot transfer.
To simplify the analysis,   the explicit time-dependent detuning used to induce two rounds of inter-dot transfer is modulated as
\begin{align}
\epsilon^{}_{2\circlearrowleft}(t)= V^{}_{e} +V^{}_{a}\sum^{}_{n=1,2}\Upsilon^{}_{n} (t)\ ,
\end{align}  with
\begin{align}
\Upsilon^{}_{n} (t)= J^{}_{n}(t-w^{}_{0})
   \tilde{H}( t-\varsigma^{}_{n,1}) [1-\tilde{H}(t-\varsigma^{\prime}_{ n,1})]+ J^{}_{n}
  (z^{}_{n}+w^{}_{0}-t)  \tilde{H}( t-\varsigma^{}_{n,2}) [1-\tilde{H}(t-\varsigma^{\prime}_{ n,2})]\ .
   \end{align}   Specifically,  we define the following functions and parameters  herein :    $J^{}_{1} (t) =[\tanh(\pi t/w^{}_{0})+1]/2$,  $J^{}_{2}(t)=\{\tanh[\pi (t-k_1)/w^{}_{0}]+1\}/2$, $k^{}_{1}=4w_{0}+t_{in,1}+t^{}_{out,1}$,  $ k^{}_{2}=8w^{}_{0}+\sum^{}_{n=1,2}(t^{}_{in,n}+t^{}_{out,n}) $,   $\varsigma^{}_{1,1 } = 0$, $\varsigma^{}_{2,1 }=k^{}_{1}$,  $\varsigma^{\prime}_{n,1 }=\varsigma^{}_{n,2}=z^{}_{n}$,  and   $\varsigma^{\prime}_{n,2 }=k^{}_{n}$, in which   $ z^{}_{1} =2w^{}_{0}+t^{}_{in,1}$,   $z^{}_{2} =k^{}_{1}+2w^{}_{0}+t^{}_{in,2}$,  $t^{}_{in/out,n}$  indicate the respective idling times within the left and right QDs ,    and   $\tilde{H}(t)$  represents  the Heaviside function.   The time evolution  of  $\epsilon^{}_{ 2\circlearrowleft} (t)$ is exhibited in Fig.~5(a) of the main text.
 In addition, the difference between the baseline value ($V^{}_{e}$) and the plateau value ($V^{}_{c}=V^{}_{e}+V^{}_{a}$) of the modulated detuning is set sufficiently large to ensure high-fidelity inter-dot transfer during each unidirectional shuttling process.   By  analogy with Eq.~(\ref{Shum}) of the main text,  the connection matrix for describing  the leftward/rightward transfer in  the $n$-th round   can  then be presented as
\begin{align}
\hat{S}^{}_{l/r,n} =\left(\begin{array}{cc}
e^{i\phi^{}_{l/r,n}}_{} \cos\frac{\vartheta^{}_{s,n}}{2} & -e^{-i\varphi^{}_{l/r,n}-i\phi^{}_{l/r,n}}_{}\sin
\frac{\vartheta^{}_{s,n}}{2} \\ \\
e^{i\varphi^{}_{l/r,n}+i\phi^{}_{l/r,n}}_{} \sin
\frac{\vartheta^{}_{s,n}}{2} &
 e^{-i\phi^{}_{l/r,n}}_{}\cos\frac{\vartheta^{}_{s,n}}{2}
 \end{array}\right)\ .~\label{shut}
\end{align}
   Considering the  modulation  in the tunnel-coupling strength  given by $T^{}_{0}=\begin{cases}
    T^{}_{1} ~~~ t\leq  k^{}_{1}\\
    T^{}_{2}~~~t >k^{}_{1}
     \end{cases}$ ,  it can be inferred that the characteristic parameters of the connection matrix, namely  $\vartheta^{}_{s,n}$,
  $\varphi^{}_{l/r,n}$,  and  $\phi^{}_{l/r, n}$,  exhibit distinct values between the first and second transfer rounds. Specifically,
      the  distinct values  of these basic characteristic parameters,  corresponding to the    different values of  $w^{}_{0}$ and $T^{}_{n }$  ($n=1,2$), are listed in Table~\ref{tab1}.

 \begin{table}[t]
  \caption{The  characteristic  parameters  of the  connection  matrix in Eq.~(\ref{shut}), i.e., $\vartheta^{}_{s,n}$, $\varphi^{}_{l ,n}$, $\varphi^{}_{r,n}$,$\phi^{}_{l ,n}$, and $\phi^{}_{ r,n}$, with   $ V^{}_{c/e} =\pm 10$ (in a unit of $\Delta^{}_{0}$) and  different  values of    $T^{}_{n}$  and $w^{}_{0}$. Besides, in the specific  modulations,  the  direction angles of the QDs' spin quantization  are set as $\theta^{}_{l}=0.95\pi$ and $\theta^{}_{r}=0.44\pi$  and the SOI-induced rotation phase is   $\phi^{}_{\rm so}=0.11\pi$.   }
  % \label{TA}
 \centering
 \begin{tabular}{| c|c|c|c|c|c|c|c|c|c|c|c|c|}
 \hline
  ~~ $w^{}_{0}$ ~~ &~~$T^{}_{1}$~~&~~$\vartheta^{}_{s,1}$~~ & ~~$\varphi^{}_{l,1}~~$
 & ~~$\varphi^{}_{r,1}$~ ~&~~ $\phi^{}_{l,1}$~~ & ~~$\phi^{}_{r,1}$ ~~&~~$T^{}_{2}$~~&~~$\vartheta^{}_{s,2}$~~ &~~ $\varphi^{}_{l,2}$
 ~~& ~~$\varphi^{}_{r,2}$~ ~&~ ~$\phi^{}_{l,2}$~~ &~ ~$\phi^{}_{r,2}$~\\ \hline
~~ $5t^{}_{0}$ ~~& $2.00$ & $0.28\pi$&$1.00\pi$& $1.76\pi$ &$0.08\pi$&$0.32\pi$&$2.20$& $0.25\pi$ &$1.05\pi$ & $1.80\pi$ &$0.04\pi$ &$0.28\pi$ \\ \hline
~~  $6t^{}_{0}$ ~~ &$1.80$ & $0.27\pi$ & $1.05\pi$ & $1.80\pi$ & $0.04\pi$ & $0.28\pi$&$2.00$& $0.24\pi$ & $1.09\pi$ &$1.84\pi$ &$0.00$& $0.24\pi$\\
 \hline
 \end{tabular}
 \label{tab1}
 \end{table}

By incorporating the effect of short stays in the QDs, which  corresponds to  the Z-gate operations,  the overall evolution matrix for describing the two rounds of   inter-dot shuttling protocol is given by   Eq.~(\ref{UST}) of the main text. To facilitate the analysis, the operation matrix can be reformulated as
\begin{align}
\mathbf{U}^{}_{2\circlearrowleft} =\hat{Z}^{}_{} (t^{}_{out,2}) \widehat{\rm Tw}^{}_{2} (t^{}_{in,2}) \hat{Z}^{}_{} (t^{}_{out,1}) \widehat{\rm Tw}^{}_{1}(t^{}_{in, 1})\ ,
\label{TYU}
\end{align}
with $\widehat{\rm Tw}^{}_{n} (t^{}_{in,n}) =\hat{S}^{}_{r,n} \hat{Z}^{}_{} (t^{}_{in,n}) \hat{S}^{}_{l,n}$. By analogy with Eq.~(\ref{mat}), it is found that
\begin{align}
\widehat{\rm Tw}^{}_{n} (t^{}_{in,n})  = \left(\begin{array}{cc}
e^{i\phi^{}_{l,n}}_{} {\rm M}^{ }_{11,n} & e^{-i\phi^{}_{l,n}}_{} {\rm M}^{}_{12,n}\\
e^{i\phi^{}_{l,n}}_{} {\rm M}^{ }_{21,n} & e^{-i\phi^{}_{l,n}}_{} {\rm M}^{}_{22,n}
\end{array}\right)\ ,
\label{Twn}
\end{align}
in which the matrix elements  are evaluated as
\begin{align}
 {\rm  M}^{}_{11,n}= &e^{i \varphi^{}_{\delta,n} }_{} [\cos \zeta^{}_{n}
   \cos\vartheta^{}_{s,n}+i\sin \zeta^{}_{n}]\nonumber\\
    {\rm M}^{}_{12,n } =&-e^{-i\varphi^{}_{0,n}}_{} \sin\vartheta^{}_{s,n} \cos \zeta^{}_{n}\ .
\end{align}
with $\zeta^{}_{n}=
   g^{}_{0} t^{}_{in,n} +\phi^{}_{d,n }$,  $\varphi^{}_{\delta,n} =  (\varphi^{}_{l,n}-\varphi^{}_{r,n})/2$ and
 $\phi^{}_{d,n}=\phi^{}_{r,n}-(\varphi^{}_{l,n}-\varphi^{}_{r,n})/2$. Substituting Eq.~(\ref{Twn}) into Eq.~(\ref{TYU}) and by introducing two auxiliaries $A^{}_{n} (t^{}_{in,n})=
 \cos\zeta^{}_{n}\cos\vartheta^{}_{s,n} +i \sin \zeta^{}_{n}$,  the evolution matrix can be derived as
\begin{align}
\mathbf{U}^{}_{2\circlearrowleft} = \left(\begin{array}{cc}
 {\rm \Sigma T}^{}_{1} &  -{\rm \Sigma T}^{}_{2}\\
 {\rm \Sigma T}^{\ast}_{2} & {\rm \Sigma T}^{\ast}_{1}
\end{array}\right)\ ,
\label{UM}
\end{align} in which the  matrix elements are   calculated as
\begin{align}
{\rm \Sigma T}^{}_{1} =&
 e^{ig^{}_{0} t^{}_{out,2}+iQ^{}_{1}}_{} [e^{ig^{}_{0}t_{out,1} +i\lambda}_{}\cos\omega^{}_{1}  \cos\omega^{}_{2} -e^{-i g^{}_{0}t_{out,1} -i\lambda}_{} \sin\omega^{}_{1} \sin\omega^{}_{2} ]
 \nonumber\\
{\rm \Sigma T}^{}_{2} =  &e^{i g^{}_{0}t^{}_{out,2}-iQ_{2}}_{} [e^{ig^{}_{0} t^{}_{out,1}+i\lambda}_{} \sin\omega^{}_{1}\cos\omega^{}_{2} +e^{-i g^{}_{0} t^{}_{out,1}-i\lambda}_{} \cos\omega^{}_{1} \sin\omega^{}_{2}]\ .
\label{bsa}
\end{align}
Here  $\omega^{} _{n=1,2} =\arccos\left[|A^{}_{n} (t^{}_{in,n})|\right]$,   $Q^{}_{1/2}=  \eta^{}_{s/c} \mp \varphi^{}_{21,c/s} +\phi^{}_{l,1}$, and  $\lambda= \eta^{}_{s}+\phi^{}_{l,2}-\phi^{}_{12,c}$, with   $\eta^{}_{s/c}= [\arg( A^{}_{1}) \pm \arg( A^{}_{2})]/2$ and
$\varphi^{}_{nn^{\prime}_{},c/s} = (\varphi^{}_{r,n} \mp  \varphi^{}_{l,n^{\prime}_{}})/2$.

Evidently,   the vanishing of the  diagonal  elements in Eq.~(\ref{UM}), i.e., $\Sigma^{}_{} T^{}_{1}=0$,  is  the precondition for implementing
the single-qubit X or Y gate. Using the concrete form of ${\rm \Sigma^{}_{} T}^{}_{1}$ in Eq.~(\ref{bsa}), this precondition can be reformulated into
\begin{align}
\cos (g^{}_{0}t^{}_{out,1} +\lambda )[\cos\omega^{}_{1}  \cos\omega^{}_{2} -\sin\omega^{}_{1} \sin\omega^{}_{2}]+i \sin(g^{}_{0}t^{}_{out,1} +\lambda )  [\cos\omega^{}_{1}  \cos\omega^{}_{2} +\sin\omega^{}_{1} \sin\omega^{}_{2}]=0\ .
\label{nsb}
\end{align}
Due to    the   fact of  $   \omega^{}_{j=1,2}   \in (0,\pi/2]$, it can be inferred that Eq.~(\ref{nsb})  holds only under the following two conditions:
\begin{align}
 \sin (g^{}_{0}t^{}_{out,1} +\lambda) =0
 \label{cov1}
 \end{align}  and
   \begin{align}\cos\omega^{}_{1}\cos\omega^{}_{2} -\sin\omega^{}_{1}\sin\omega^{}_{2}=0 ~\Rightarrow  \Omega^{}_{ }  =\pi/2\ ,
\label{cons}
\end{align}
with $\Omega^{}_{} =\omega^{}_{1}+\omega^{}_{2}$.
Under this circumstance,     the  operation matrix in Eq.~(\ref{UM}) can be further  simplified into
\begin{align}
\mathbf{U}^{}_{2\circlearrowleft}=\left(\begin{array}{cc}
0 & -e^{ig^{}_{0}t^{}_{out,2}-iQ^{}_{2}}_{} \\
e^{-ig^{}_{0} t^{}_{out,2}+iQ^{}_{2}}_{} & 0\end{array}\right)\ .
\label{usc}
\end{align}
Then, it follows that $\mathbf{U}^{}_{2\circlearrowleft}$  in Eq.~(\ref{usc}) can be  equivalent to the Pauli-X  matrix $  {\rm X}= \left(\begin{array}{cc}
0&1 \\
1& 0
\end{array}\right)$    when  $|\sin( g^{}_{0} t^{}_{out,2}-Q^{}_{2})|= 1 $.  Analogously, the Pauli-Y matrix    $ {\rm Y}=\left(\begin{array}{cc}
0&-i\\
i&0
\end{array}\right)$ can also be realized  if $|\sin( g^{}_{0} t^{}_{out,2}-Q^{}_{2})|= 0$.

Interestingly,  when the idling time $t^{}_{out,1}$  meets exclusively the constraint defined in Eq.~(\ref{cov1}),     $\mathbf{U}^{}_{2\circlearrowleft}$ can be  reduced  to  Eq.~(\ref{us2}) of the main text
\begin{align}
\mathbf{U}^{}_{2\circlearrowleft}=\left(\begin{array}{cc}
\cos\Omega e^{ig^{}_{0}t^{}_{out,2} +iQ^{}_{1}}_{}   &
-\sin\Omega e^{ig^{}_{0}t^{}_{out,2} -iQ^{}_{2}}_{} \\
\sin\Omega  e^{-ig^{}_{0}t^{}_{out,2} +iQ^{}_{2}}_{} &  \cos\Omega e^{-ig ^{}_{0}t^{}_{out,2} -iQ^{}_{1}}_{}
\end{array}\right)\ .
\end{align}
For $\Omega= (2\pm 1)\pi/4$, and with $t^{}_{r,2}$ fulfilling    $\cos(g^{}_{0} t^{}_{out,2}+Q^{}_{1}) =0$, it is  found that   the operation matrix  can be mapped to a generalized Hadamard gate
\begin{align}
{\rm H}=\frac{1}{\sqrt{2}} \left(\begin{array}{cc}
1 & e^{i\varphi^{}_{0}}_{} \\
 e^{-i\varphi^{}_{0}}_{}& -1
\end{array}\right)\ .\label{Had}
\end{align}
In contrast to a conventional Hadamard gate, the gate modulation involves an extra phase shift $ \varphi^{}_{0}$  given by   $\varphi^{}_{0}= \tilde{\varphi}^{}_{} \mp \pi/2$, with $\tilde{\varphi}=\pi/2-Q_{1}-Q^{}_{2}$ and $\Omega= (2\pm 1)\pi/4$.

%%%%%%%%%%%%%%%%%%%%%%%%%%%%%%%%%%%%%%%%%%%%%%%%%
%%%%%%%%%%%%%%%%%%%%%%%%%%%%%%%%%%%%%%%%%%%%%%%%%

% in the regulation of the exchange interaction strength.

%%%%%%%%%%%%%%%%%%%%%%%%%%%%%%%%%%%%%%%%%%%%%%%%%
%%%%%%%%%%%%%%%%%%%%%%%%%%%%%%%%%%%%%%%%%%%%%%%%%

%%%%%%%%%%%%%%%%%%%%%%%%%%%%%%%%%%%%%%%%%%%%%%%%%
%%%%%%%%%%%%%%%%%%%%%%%%%%%%%%%%%%%%%%%%%%%%%%%%%

%%%%%%%%%%%%%%%%%%%%%%%%%%%%%%%%%%%%%%%%%%%%%%%%%
%%%%%%%%%%%%%%%%%%%%%%%%%%%%%%%%%%%%%%%%%%%%%%%%%

%%%%%%%%%%%%%%%%%%%%%%%%%%%%%%%%%%%%%%%%%%%%%%%%%
%%%%%%%%%%%%%%%%%%%%%%%%%%%%%%%%%%%%%%%%%%%%%%%%%

%%%%%%%%%%%%%%%%%%%%%%%%%%%%%%%%%%%%%%%%%%%%%%%%%
%%%%%%%%%%%%%%%%%%%%%%%%%%%%%%%%%%%%%%%%%%%%%%%%%

%%%%%%%%%%%%%%%%%%%%%%%%%%%%%%%%%%%%%%%%%%%%%%%%%
%%%%%%%%%%%%%%%%%%%%%%%%%%%%%%%%%%%%%%%%%%%%%%%%%

 %%%%%%%%%%%%%%%%%%%%%%%%%%%%%%%%%%%%%%%%%%%%%%%%%
%%%%%%%%%%%%%%%%%%%%%%%%%%%%%%%%%%%%%%%%%%%%%%%%%We thank for useful discussion with D. Loss, J-Y Wang, P. Zhang, and  L.  Kouwenhoven.

%%%%%%%%%%%%%%%%%%%%%%%%%%%%%%%%%%%%%%%%%%%%%%%%%%%%%%%%%%%
%%%%%%%%%%%%%%%%%%%%%%%%%%%%%%%%%%%%%%%%%%%%%%%%%%%%%%%%%%%

\end{widetext}

%%%%%%%%%%%%%%%%%%%%%%%%%%%%%%%%%%%%%%%%%%%%%%%%%
%%%%%%%%%%%%%%%%%%%%%%%%%%%%%%%%%%%%%%%%%%%%%%%%%

\end{document}